\documentclass[journal]{IEEEtran}

\usepackage{glossaries, soul, graphicx, caption, subfigure, multirow, multicol, algorithm, algpseudocode, xcolor, tikz, pgfplots, fixltx2e, booktabs, hyperref, titlesec, dingbat, lipsum, makecell, xcolor}

\newacronym{rdo}{RDO}{Rate-Distortion Optimization}
\newacronym{rdcost}{RD-cost}{Rate-Distortion cost}
\newacronym{jvet}{JVET}{Joint Video Exploration Team}
\newacronym{vtm10}{VTM10}{VVC Test Model 10}
\newacronym{vtm15}{VTM15}{VVC Test Model 15}
\newacronym{vtm}{VTM}{VVC Test Model}
\newacronym{hevc}{HEVC}{High Efficiency Video Coding}
\newacronym{bdrate}{BD-rate}{Bjontegaard Delta-Rate}
\newacronym{cnn}{CNN}{Convolutional Neural Network}
\newacronym{ctc}{CTC}{Common Test Condition}
\newacronym{ragop32}{RAGOP32}{RandomAccess Group Of Picture 32}
\newacronym{ra}{RA}{RandomAccess}
\newacronym{ld}{LD}{Low-Delay}
\newacronym{ai}{AI}{All-Intra}
\newacronym{vvc}{VVC}{Versatile Video Coding}
\newacronym{hdr}{HDR}{High Definition Range}
\newacronym{vr}{VR}{Virtual Reality}
\newacronym{qtmt}{QTMT}{quadtree with nested multi-type tree}
\newacronym{ctu}{CTU}{Coding Tree Unit}
\newacronym{ns}{NS}{No Split}
\newacronym{cu}{CU}{Coding Unit}
\newacronym{qt}{QT}{Quaternary Tree}
\newacronym{bt}{BT}{Binary Tree}
\newacronym{mt}{MT}{Multi-type Tree}
\newacronym{tt}{TT}{Ternary-type Tree}
\newacronym{hbt}{HBT}{Horizontal Binary Tree}
\newacronym{vbt}{VBT}{Vertical Binary Tree}
\newacronym{vtt}{VTT}{Vertical Ternary Tree}
\newacronym{htt}{HTT}{Horizontal Ternary Tree}
\newacronym{rf}{RF}{Random Forest}
\newacronym{mltcnn}{MLT-CNN}{multi-level tree CNN}
\newacronym{uhd}{UHD}{Ultra-High Definition}
\newacronym{msecnn}{MSE-CNN}{Multi-Stage Exit CNN}
\newacronym{dt}{DT}{Decision Tree}
\newacronym{lgbm}{LGBM}{Light Gradient Boosting Machine}
\newacronym{av1}{AV1}{AOMedia Video 1}
\newacronym{jem}{JEM}{Joint Exploration Model}
\newacronym{avc}{AVC}{Advanced Video Coding}
\newacronym{svm}{SVM}{Support Vector Machine}
\newacronym{satd}{SATD}{Sum of Absolute Transform Differences}
\newacronym{sad}{SAD}{Sum of Absolute Differences}
\newacronym{qtbt}{QTBT}{quadtree plus binary tree}
\newacronym{qtdepthmap}{QTdepthMap}{Quad Tree depth map}
\newacronym{fcn}{FCN}{Fully Convolutional Network}
\newacronym{msmvfcnn}{MS-MVF-CNN}{Multi-Scale Motion Vector Field CNN}
\newacronym{msmvf}{MS-MVF}{Multi-Scale Motion Vector Field}
\newacronym{hm}{HM}{HEVC Test Model}

\newacronym{isp}{ISP}{Intra SubPartition}
\newacronym{bcnn}{B-CNN}{Branch Convolutional Neural Network}
\newacronym{mbmpcnn}{MBMP-CNN}{Multi-Branch Multi-Pooling CNN}
\newacronym{qp}{QP}{Quantization Parameter}
\newacronym{cbam}{CBAM}{Convolutional block attention module}
\newacronym{mae}{MAE}{Mean Absolute Error}

\newacronym{ts}{TS}{Time Saving}

\newacronym{flops}{FLOPs}{floating point operations}

\newacronym{tp}{TP}{True Positive}

\newacronym{fn}{FN}{False Negative}

\newacronym{tn}{TN}{True Negative}

\newacronym{fp}{FP}{False Positive}

\usepackage[space]{cite}

\titlespacing*{\subsubsection} {0pt}{3.5ex plus 3ex minus .2ex}{1ex plus .2ex}

\titlespacing*{\paragraph} {0pt}{1.5ex plus 1.5ex minus .2ex}{1ex plus .2ex}

\ifCLASSINFOpdf

\else

\fi

\begin{document}

\title{CNN-based Prediction of Partition Path for VVC Fast Inter Partitioning Using Motion Fields}

\author{Yiqun Liu,~\IEEEmembership{Student Member,~IEEE,}
        Marc Riviere,~\IEEEmembership{Fellow,~IEEE,}
        Thomas Guionnet,~\IEEEmembership{Fellow,~IEEE,}
        Aline Roumy,~\IEEEmembership{Fellow,~IEEE,}
        and Christine Guillemot,~\IEEEmembership{Fellow,~IEEE}% <-this % stops a space

\thanks{Manuscript received October 10, 2023}% <-this % stops a space
\thanks{Yiqun Liu, Aline Roumy and Christine Guillemot are with the INRIA (Institut National de Recherche en Informatique et en Automatique) Rennes Bretagne Atlantique, Rennes, France (e-mail: yiqun.liu@irisa.fr; aline.roumy@inria.fr; christine.Guillemot@inria.fr).}% <-this % stops a space
\thanks{Yiqun Liu, Thomas Guionnet and Marc Riviere are with the ATEME company Rennes Bretagne Atlantique, Rennes, France (e-mail: y.liu@ateme.com; m.riviere@ateme.com; t.guionnet@ateme.com).}}

\markboth{Paper under review for Transactions on Image Processing}%
{Shell \MakeLowercase{\textit{et al.}}: VVC Fast Inter Partitioning by Multi-Scale Motion Field Based CNN}

\maketitle

\begin{abstract}
The \gls{vvc} standard has been recently finalized by the \gls{jvet}. Compared to the \gls{hevc} standard, \gls{vvc} offers about 50\% compression efficiency gain, in terms of \gls{bdrate}, at the cost of a 10-fold increase in encoding complexity. In this paper, we propose a method based on \gls{cnn} to speed up the inter partitioning process in \gls{vvc}. Firstly, a novel representation for the \gls{qtmt} partition is introduced, derived from the partition path. Secondly, we develop a U-Net-based \gls{cnn} taking a multi-scale motion vector field as input at the \gls{ctu} level. The purpose of \gls{cnn} inference is to predict the optimal partition path during the \gls{rdo} process. To achieve this, we divide \gls{ctu} into grids and predict the \gls{qt} depth and \gls{mt} split decisions for each cell of the grid. Thirdly, an efficient partition pruning algorithm is introduced to employ the \gls{cnn} predictions at each partitioning level to skip \gls{rdo} evaluations of unnecessary partition paths. Finally, an adaptive threshold selection scheme is designed, making the trade-off between complexity and efficiency scalable. Experiments show that the proposed method can achieve acceleration ranging from 16.5\% to 60.2\% under the \gls{ragop32} configuration with a reasonable efficiency drop ranging from 0.44\% to 4.59\% in terms of \gls{bdrate}, which surpasses other state-of-the-art solutions. Additionally, our method stands out as one of the lightest approaches in the field, which ensures its applicability to other encoders.
\end{abstract}


\IEEEpeerreviewmaketitle

\section{Introduction}
% \textcolor{red}{Paragraph 1 introducing the VVC codec and the need for accelerating inter coding by fast partitioning method}
\IEEEPARstart    {A}{ccording} to \cite{cisco}, global internet traffic has increased substantially, primarily due to the growing video usage, which now accounts for 65\% of internet traffic. In addition, the rapid development of \gls{uhd} and \gls{vr} makes it critical to design more efficient video compression codecs. For this purpose, the latest video coding standard \gls{vvc} has been finalized in 2020. In comparison to its predecessor, \gls{hevc}, its efficiency of inter coding is boosted by about 50\% in terms of \gls{bdrate} at the cost of 10 times higher complexity~\cite{bross2021overview}. The substantial complexity of \gls{vvc} impedes its direct implementation in real-time applications such as TV broadcasting. Apart from multiple newly added inter coding tools~\cite{wang2016adaptive, li2017efficient, alshin2010bi}, a novel partition structure introduced in \gls{vvc}, called \gls{qtmt}~\cite{huang2021block}, is the main contributor to this complexity surge. In particular, it has been observed in \cite{complexity} that the \gls{vtm} encoder, which is an implementation of the \gls{vvc} codec, dedicates 97\% of its encoding time to searching for the optimal partition. Consequently, fast partitioning methods emerge as the most promising approaches to speed up the whole \gls{vvc} encoding process.

\subsection{Partitioning Acceleration for \gls{vvc}}

\subsubsection{Fast Intra Partitioning Methods}

% \textcolor{red}{Paragraph 2 categorizing intra fast partitioning methods}
Numerous works achieve an important acceleration of intra-frame partitioning in the \gls{ai} encoding configuration. These approaches fall primarily into two categories: heuristic-based methods and machine learning-based methods.

% \textcolor{red}{Paragraph 3 presenting intra heuristic-based methods}
Some heuristic-based methods are built upon pixel-wise statistics, such as gradients~\cite{fan2020fast, cui2020gradient, chen2019vcip} and variances~\cite{fan2020fast, chen2019vcip}. To simplify the partitioning process, other heuristic methods reuse some data generated during the encoding process, such as \gls{rdcost} of \gls{cu} encoding ~\cite{lei2019look}, coding tool decisions~\cite{saldanha2020fast}, best split type, and the intra mode of sub-\gls{cu}s~\cite{fu2019icme}.

% \textcolor{red}{Paragraph 4 presenting intra machine learning methods}
Machine learning-based methods utilize \gls{cnn} or \gls{dt} models to expedite intra partitioning. In \cite{galpin2019cnn, tissier2022machine, wu2022hg}, a \gls{cnn} model is trained to predict the split boundaries inside \gls{ctu} partitions. In \cite{Feng2023cnn}, Feng \emph{et al.} propose a fast partitioning method by predicting a \gls{qt} depth map, multiple \gls{mt} depth maps, and multiple \gls{mt} direction maps with \gls{cnn}. Regarding the \gls{dt}-based approach, various \gls{lgbm} classifiers are separately trained for different \gls{cu} sizes to predict the possible splits, as demonstrated in \cite{saldanha2021configurable}.

\subsubsection{Fast Inter Partitioning Methods}\label{1.1.2}

% \textcolor{red}{Paragraph 5 explaining the challenge for inter fast partitioning and its importance}
Fewer contributions of fast inter partitioning methods have been proposed for \gls{vvc}. Due to the fact that the inter coding consists of predicting pixels of current frame depending on previously encoded reference frames, encoding errors resulting from the use of fast coding methods are propagated and accumulated between frames. Therefore, the acceleration of inter partitioning is a more challenging task compared to that of intra partitioning. Nevertheless, the acceleration of inter-frame coding is key to speeding up the overall encoding process, especially in \gls{ra} and \gls{ld} configurations. These configurations are employed more widely than the \gls{ai} configuration in scenarios such as broadcasting and streaming.

% \textcolor{red}{Paragraph 6 presenting heuristic methods for inter fast partitioning and their drawbacks}
Generally, fast partitioning approaches aim to reduce the search space of potential partitions. Therefore, accurately predicting the subset of partitions is of crucial importance. Heuristic methods proposed for fast intra partitioning of \gls{vvc}~\cite{fan2020fast, fu2019icme} heavily depend on handcrafted features to determine whether to check a partition. These methods are fast and simple to implement but lack accuracy for two reasons. Firstly, the features are computed locally on the \gls{cu} and/or sub-\gls{cu}s, which fails to provide a synthesized view of the entire \gls{ctu}. Secondly, these features, including variances, gradients, and coding information, are low dimensional and do not adequately capture the complexity of \gls{ctu}.

% \textcolor{red}{Paragraph 7 presenting rf/dt based methods and their limitations}
One approach to improve the accuracy of partition prediction involves increasing the dimension of the extracted features. This is the case with the methods based on \gls{rf}~\cite{amestoy2019tunable} or \gls{dt}~\cite{kulupana2021fast}, which use over 20 features from a given \gls{cu} and its sub-\gls{cu}s. As a result, decisions made by these methods remain confined to the local context of \gls{cu}, without considering the entirety of \gls{ctu}. Rather than relying on local information, a more effective selection of subsets of partitions should be based on global features computed on the entire \gls{ctu}. This can be accomplished through the utilization of \gls{cnn}-based methods.

% \textcolor{red}{Paragraph 8 introducing the CNN based methods for partial acceleration} 
Several approaches~\cite{pan2021cnn, yeo2021cnn, liu2022light} use \gls{cnn} to partially accelerate the partition search process. For instance, in \cite{pan2021cnn}, Pan \textit{et al.} propose a multi-branch \gls{cnn} to perform a binary classification of the “Partition” or “Non-partition” at the \gls{cu} level. In \cite{yeo2021cnn}, the split type at the \gls{ctu} level is predicted, whereas the partitions of its sub-\gls{cu}s are not determined. Liu \textit{et al.} in \cite{liu2022light} employ a \gls{cnn} to estimate an 8x8 grid map of \gls{qt} depth, which is used to discard a portion of the \gls{mt} splits. These methods cover only a part of the partition search space, while the partition search is conducted exhaustively on the remainder. These methods could be referred as partial partitioning acceleration methods by \gls{cnn}.

% \textcolor{red}{Paragraph 9 introducing one CNN based method for complete acceleration}
A complete partitioning acceleration of inter coding by \gls{cnn} is proposed in \cite{tissier2022machine}. A vector containing probabilities of the existence of split boundaries in the partition is predicted similarly to \cite{tissier2023machine}. This method is fast in the sense that a single vector is computed for each \gls{ctu}. Nevertheless, it is observed in \cite{tissier2023machine}, that the predictions are more accurate at higher levels of the partitioning tree. Hence, they propose improving the decisions by adding 16 trained \gls{dt}s to process the \gls{cnn} output, introducing additional complexity to the method.

\subsection{Proposed Method}\label{1.1.3}

% \textcolor{red}{Paragraph 10 and 11 explaining the motivation for our proposed acceleration method}
In the \gls{mt} partitioning, both binary and ternary (with sub-\gls{cu}s of two different sizes) splits are available. Consequently, \gls{cu}s at a specific depth in the tree do not correspond to the same size and shape, introducing dependence between the \gls{mt} splits along the partition path. This dependence partly explains the decrease in partition prediction accuracy as the depth of the partitioning tree increases, as observed in recent studies \cite{tissier2023machine, tissier2022machine} presented in the previous section.

More precisely, since the size and shape of a \gls{cu} depend not only on its depth in the tree but also on consecutive \gls{mt} splits, depth alone is insufficient for defining a partition. Therefore, we propose making decisions on \gls{mt} partitioning in a hierarchical manner, considering their dependence on the partition path. We also introduce a one-shot approach for \gls{qt} partitioning which precedes \gls{mt} partitioning, since there is a one-to-one correspondence between the \gls{qt} depth and the \gls{cu} size at that particular depth.

% \textcolor{red}{Paragraph 11 providing a resume of our method}
Hence, our overall proposition involves predicting the partition path, which includes a one-shot prediction for the \gls{qt} partitioning, followed by a hierarchical prediction for the \gls{mt} partitioning. Additionally, to further improve the accuracy of partition prediction, we suggest basing the partition decision not only on pixel values and residual values but also on motion vector fields, as these fields exhibit a strong correlation with partitioning \cite{amestoy2019tunable}.

Our two main contributions are as follows:

\begin{itemize}

\item We propose a novel partition-path-based representation of the \gls{qtmt} partition at the \gls{ctu} level as a map of \gls{qt} depth plus three maps of \gls{mt} split well adapted to the sophisticated partitioning scheme in \gls{vvc}. \newline

\item We design a U-Net-based \gls{cnn} model taking multi-scale fields of motion vectors as input to effectively predict \gls{qt} depth map as well as split decisions at different \gls{mt} levels. \newline

\end{itemize}

We also have other contributions such as:

\begin{itemize}

\item We build MVF-Inter\footnote{Our dataset MVF-Inter is available at \url{https://1drv.ms/f/s!Aoi4nbmFu71Hgx9FJphdskXfgIVo?e=fXrs0o
}\label{mvf_inter}}, a large scale dataset for inter \gls{qtmt} partition
of \gls{vvc}, which could facilitate the research in this field. \newline

\item We propose a fine tuned loss function for this complex multi-branch multi-class classification problem. \newline

\item We develop a fast partition scheme effectively exploiting the prediction of a \gls{cnn} model in a way that the most possible splits are determined at each partition level. \newline

\item We design a specific threshold-based selection approach to match with the partition scheme, which allows us to realize a large range of trade-offs between complexity and compression efficiency. \newline

\end{itemize}

% \textcolor{red}{Paragraph 12 giving the outline of the paper}
The remainder of this paper is organized as follows. In Section \ref{2}, we provide an overview of \gls{qtmt} partitioning scheme in \gls{vvc}, including the concept of the partition path. The motivation and detailed description for our proposed representation of the \gls{qtmt} partition are presented in Section \ref{3}. In Section \ref{4}, the structure of the proposed \gls{cnn} model is illustrated. We give a detailed description of the partitioning acceleration algorithm in Section \ref{5}. The loss function of \gls{cnn} and the dataset generation process are described in Section \ref{6}. In section \ref{7}, the evaluation of the prediction accuracy of our \gls{cnn} model is carried out. Furthermore, we compare our result with the state-of-the-art of \gls{rf}-based methods and \gls{cnn}-based methods, respectively. The complexity analysis of our method is also included in this section. Finally Section \ref{8} concludes the paper. Our source code is available
at \url{https://github.com/Simon123123/MSMVF_VVC_TIP2023.git}.

\section{Overview of Partitioning Structure in VVC}\label{2}

\subsection{QTMT Partitioning}\label{2.1}
\begin{figure}[h]
    \centering
    \includegraphics[width=1\linewidth]{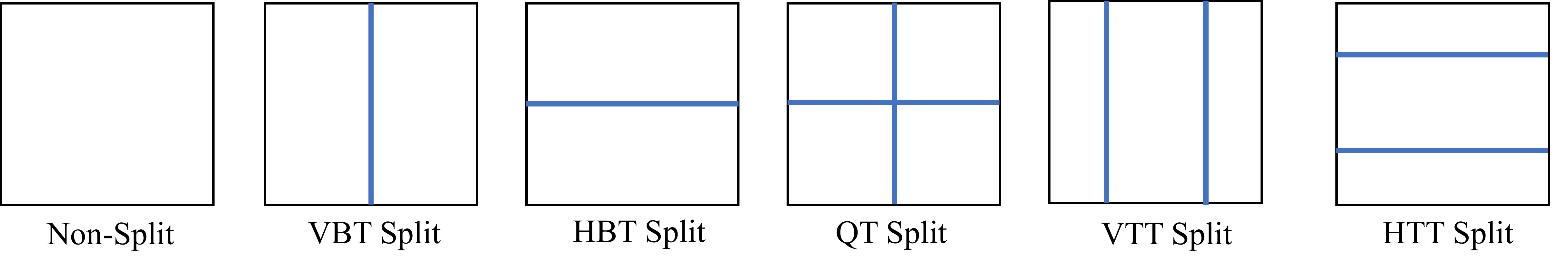}
    \caption{Different split types in VVC.}
    \label{fig:split_types}
\end{figure}

% \textcolor{red}{Paragraph 13 presenting QTMT and split types}
The partitioning of \gls{vvc} is done in a top-to-bottom manner. Starting from \gls{ctu}, the encoder applies possible split types recursively on \gls{cu} at each level of the partitioning tree, in order to find the partition which best exploits spatial and temporal redundancy. The \gls{qtmt} partitioning scheme in \gls{vvc} consists in splitting the \gls{cu} using either a \gls{qt} split or a \gls{mt} split. For the \gls{mt} split, four split types are introduced: \gls{hbt} split, \gls{vbt} split, \gls{htt} split and \gls{vtt} as illustrated in Figure \ref{fig:split_types}. 

\begin{figure}[h]
    \centering
    \includegraphics[width=0.9\linewidth]{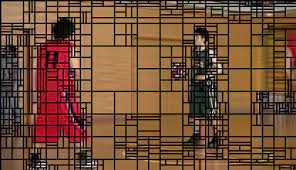}
    \caption{Example of a partitioned frame in VVC.\cite{fan2020fast}}
    \label{fig:paritioned_frame}
\end{figure}

% \textcolor{red}{Paragraph 14 presenting CU sizes in partitioning and an partitioned frame}
Based on the \gls{qtmt} partitioning scheme, dimensions of the final encoded \gls{cu} could range from 128×128 to 4×4 \cite{huang2021block} including squared and rectangular \gls{cu}s of 32 different sizes in the \gls{ra} configuration with the \gls{ctu} size of 128x128. Figure \ref{fig:paritioned_frame}, \gls{qtmt} can achieve fine partitioning adapted to the local frame texture.

\begin{figure}[h]
    \centering
    \includegraphics[width=0.9\linewidth]{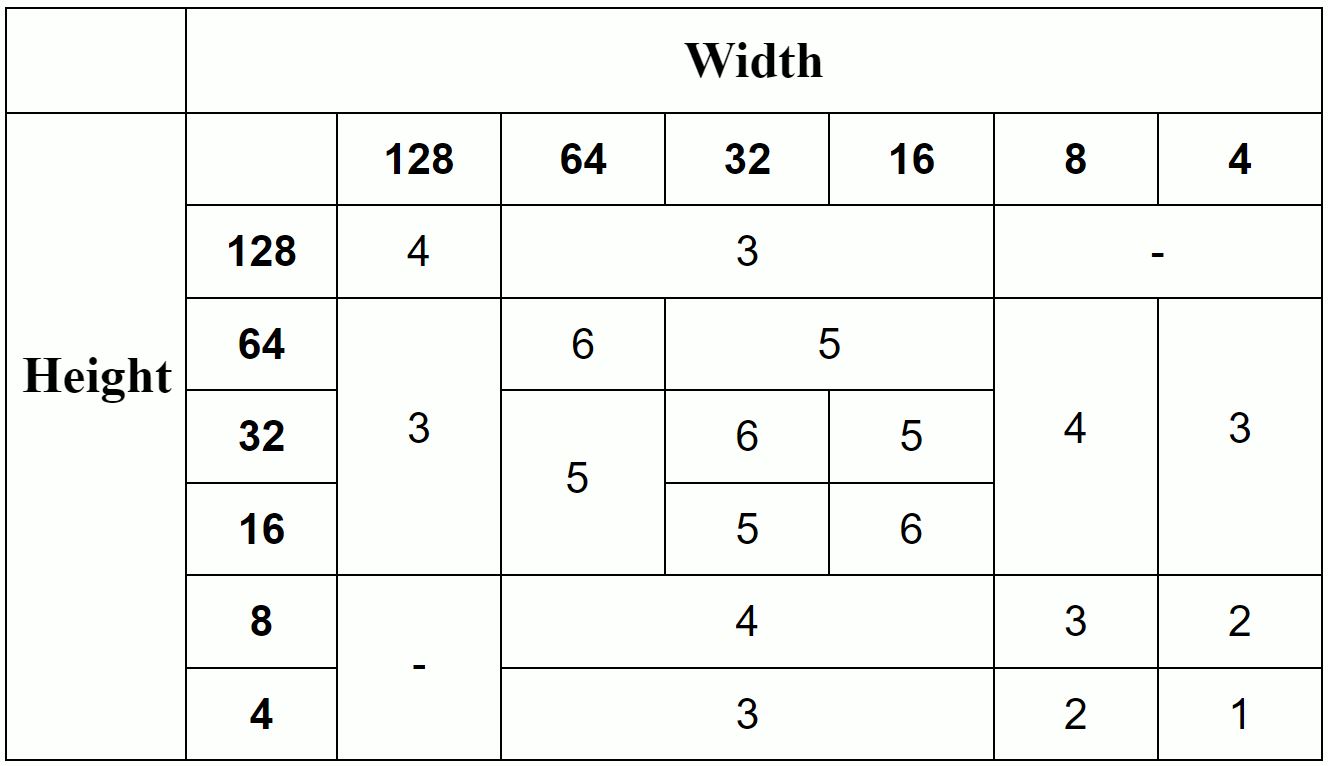}
    \caption{Number of split types per CU in VVC for Luma.}
    \label{fig:split_cu}
\end{figure}

% \textcolor{red}{Paragraph 15 presenting available split types by CU sizes and existing shortcuts}
A worth-noting characteristic of \gls{qtmt} partitioning is that the available split types depend on \gls{cu} sizes. Figure \ref{fig:split_cu} presents the number of available split types including \gls{ns} per \gls{cu} size for luma samples. The number of split options varies from 1 to 6 for different sizes of \gls{cu}, making the partition acceleration at the \gls{cu} level more complicated. Except for the \gls{cu} size restrictions on available split types, \gls{vvc} implements various shortcuts \cite{wieckowski2019fast}, including speed-up rules based on content gradients and \gls{qt} search restrictions estimated on neighboring \gls{cu}s, etc.

% \textcolor{red}{Paragraph 16 presenting the dual tree in VVC and clarifying our method is only for luma}
The partitioning is executed at \gls{ctu} level with dual tree allowing separated partitioning tree for luma and chroma. Our algorithm only accelerates the luma partitioning search using luma samples, and the resulting luma partitioning tree is then applied to the chroma components.

\subsection{Partition Path}\label{2.2}
\begin{figure}[h]
    \centering
    \includegraphics[width=1\linewidth]{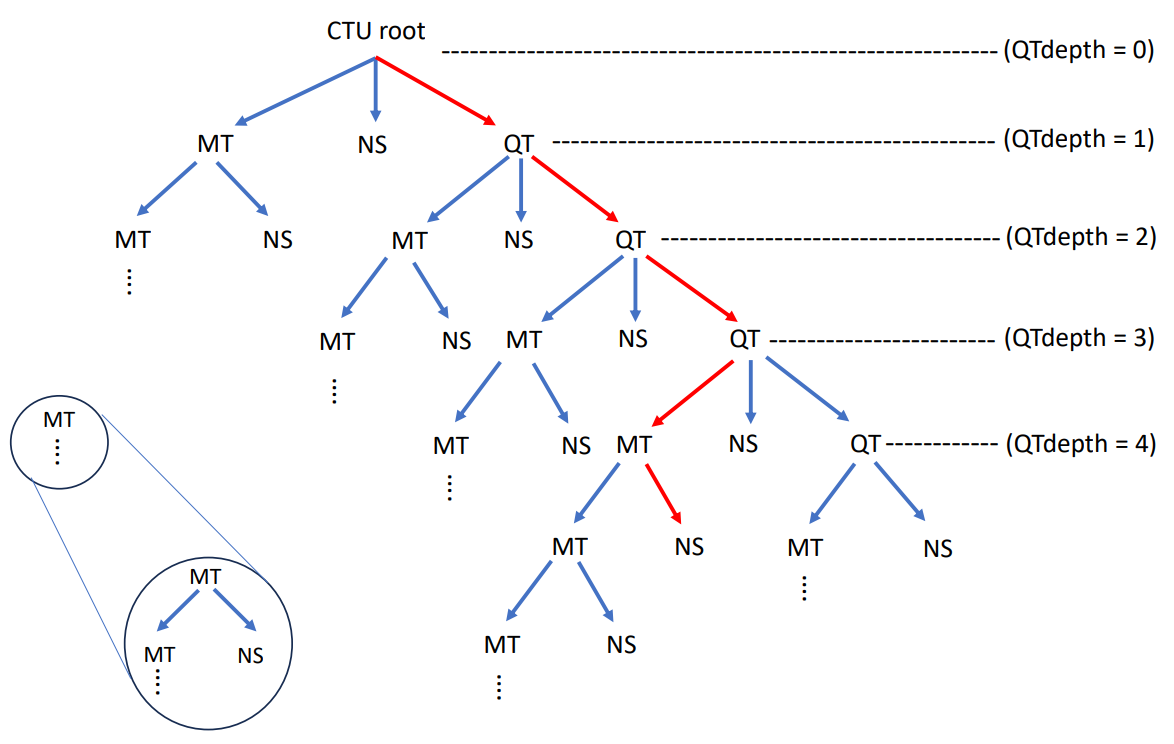}
    \caption{Example of partition paths.}
    \label{fig:qt_path}
\end{figure}

% \textcolor{red}{Paragraph 17 introducing the concept of partition path and describing (with an example) the partitioning process from the perspective of partition path}
In this work, we introduce the concept of partition path to depict the partition of a \gls{cu}. The partition path refers to the sequence of splits applied to obtain a \gls{cu} during the partitioning. In the \gls{rdo} process of partitioning, numerous partition paths included in the partition search space are checked and the optimal one leading to the final partition is selected. Figure \ref{fig:qt_path} illustrates a simplified tree representation showing all possible partition paths checked for a \gls{ctu}. The red arrows indicate the selected partition path with the lowest \gls{rdcost} \cite{rdcost}, while the blue arrows represent other paths that have been tested by \gls{rdo}, but were not selected.

% \textcolor{red}{Paragraph 18 explaining that the search for partition path is two steps}
Specifically, within the \gls{qtmt} partition, it is important to note that \gls{qt} splits are prohibited for the child nodes of a \gls{mt} split. Consequently, the search for optimal partition path in \gls{vvc} can be conceptualized as a sequential two-step decision-making process, comprising a sequence of \gls{qt} splits followed by a series of \gls{mt} splits.

\section{Novel Representation of QTMT Partition} \label{3}

% \textcolor{red}{Paragraph 19 outline of this section}
Based on the \gls{qtmt} partitioning structure and the partition path of \gls{vvc} presented in the previous section, we introduce in this section a novel representation of \gls{qtmt} partition by partition path. In \ref{3.1}, we explain the motivation for this new representation. The partition path representation is illustrated in \ref{3.2}.

\subsection{Motivation}\label{3.1}

% \textcolor{red}{Paragraph 20 presenting geometric representation in literature}
Previous partition representations at \gls{ctu} level have typically used binary vectors to depict split boundaries. In \cite{galpin2019cnn, tissier2022machine, tissier2023machine}, the authors intend to predict the split boundary of each 4x4 sub-block in \gls{ctu}. Lately, Wu \textit{et al.} improve this representation in \cite{wu2022hg} by proposing hierarchical boundaries. This adaptation is designed to better align with the \gls{qtmt} partition pattern. In this work, binary labels for split boundaries of varying lengths are predicted. Collectively, these methods provide a geometric representation of the partition.

\begin{figure}[h]
    \centering
    \includegraphics[width=0.7\linewidth]{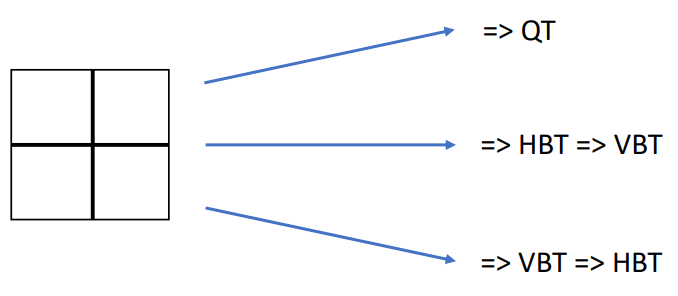}
    \caption{Possible partition paths for a final partition given by split boundaries}
    \label{fig:possible_paths}
\end{figure}

% \textcolor{red}{Paragraph 21 discussing limitations of geometric representation}
The limitations of the geometric representation mainly lie in two aspects. Firstly, it is an implicit representation of the partitioning process, requiring conversions from boundary vectors to split decisions. In the case of \cite{galpin2019cnn}, conversions are carried out by computing the average probability at the location of the specific split.
\cite{tissier2022machine} and \cite{tissier2023machine} convert boundary vectors to split decisions by \gls{dt} models separately trained for different \gls{cu} sizes. Secondly, different partition paths could be deduced from a particular partition presented in a geometric way. For example, as demonstrated in Figure \ref{fig:possible_paths}, the partition defined by the split boundaries can lead to three distinct partition paths. These partition paths correspond to different coding performances and are individually tested in the \gls{rdo} process. This multiplicity of partition paths of the geometric representation limits the acceleration potential of the method.

% \textcolor{red}{Paragraph 22 describing how our representation address these limitations}
To address the above limitations, we introduced a novel representation based on the partition path. Our representation comprises the \gls{qt} depth map and the \gls{mt} split maps. Firstly, the split decisions at each depth can be directly deduced from either the \gls{qt} depth map or the split map. This eliminates the need for decision trees, reducing method overhead, and simplifying implementation. Secondly, it corresponds to a unique partition path, maximizing the potential for complexity reduction.

\subsection{QT Depth Map and MT Split Maps}\label{3.2}

\begin{figure*}
    \centering
    \includegraphics[width=0.92\linewidth]{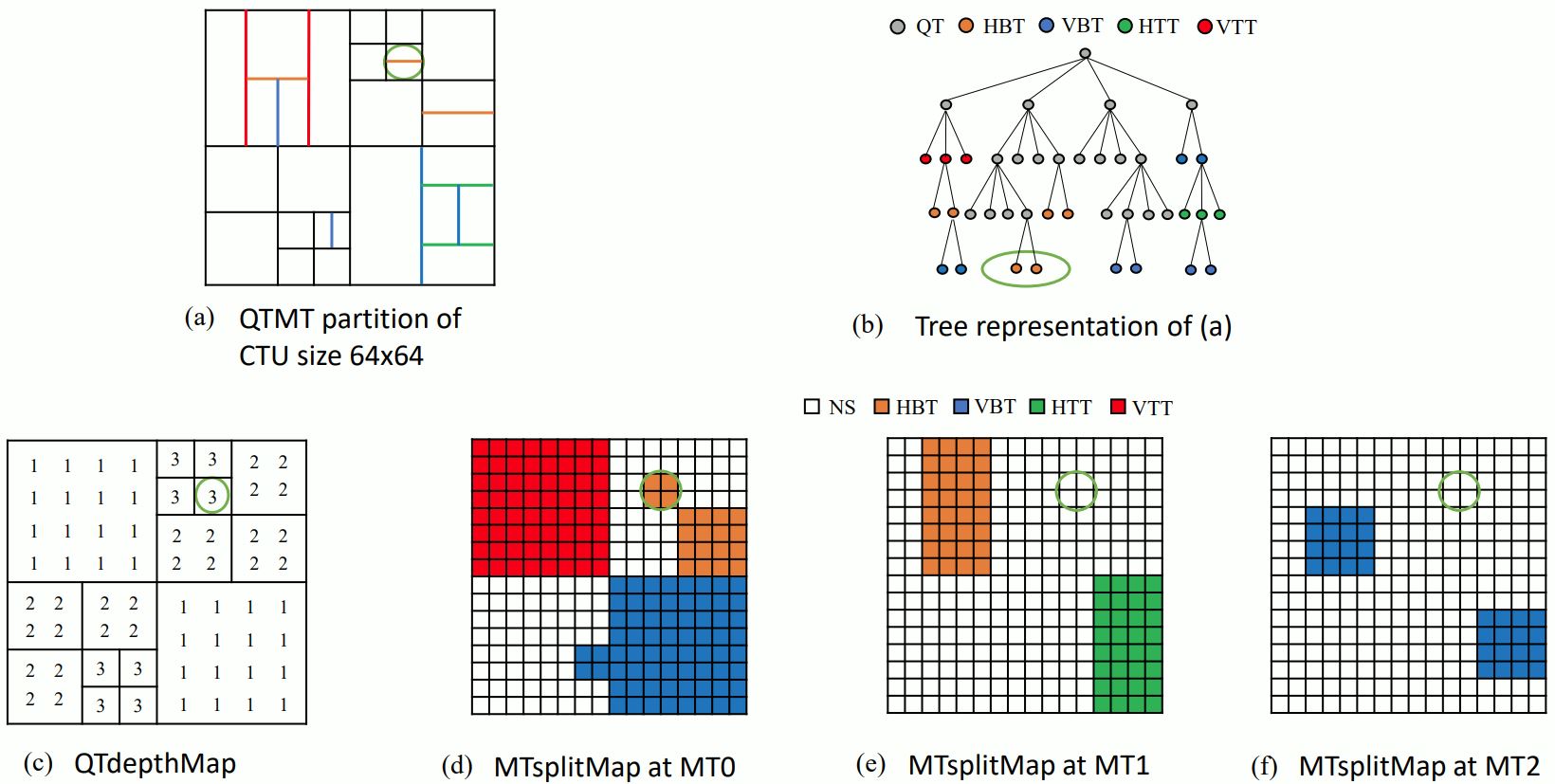}
    \caption{Example of QTMT partition, tree representation, QTdepthMap and MTsplitMaps of CTU size of 64x64}
    \label{fig:qtmt_map}
\end{figure*}

% \textcolor{red}{Paragraph 23 giving details on the form of our representation}
Considering that the maximum number of \gls{qt} splits and \gls{mt} splits is typically set to 4 and 3 in \gls{vtm}, any partition can be effectively described by a \gls{qt} depth map (\emph{i.e.} QTdepthMap) along with three \gls{mt} split maps (\emph{i.e.} MTsplitMap) in sequence. Each element within QTdepthMap and MTsplitMap corresponds to an 8x8 and 4x4 area, which aligns with the dimensions of the smallest sub-\gls{cu}s for the \gls{qt} split and the \gls{mt} split in \gls{vtm}.

% \textcolor{red}{Paragraph 24 providing an illustration of our representation}
A detailed example of our partition representation is shown in Figure \ref{fig:qtmt_map}. To keep it simple and without loss of generality, we represent this example for a \gls{ctu} size of 64×64. In this figure, (a) shows an instance of \gls{qtmt} partition with its corresponding tree representation shown in (b). (c)-(f) illustrate the QTdepthMap and MTsplitMaps generated from this partition. Given that the \gls{ctu} size in this example is 64x64, the sizes of QTdepthMap and MTsplitMap are 8x8 and 16x16, respectively. The QTdepthMap in (c) consists of QT depth values ranging from 0 to 4, while each element in MTsplitMap in (d)-(f) represents the split decision among five options: \gls{ns}, \gls{hbt}, \gls{vbt}, \gls{htt} and \gls{vtt}. This representation depicts a distinct partition path for every \gls{cu} within the partition. To provide an example, consider the \gls{cu} highlighted in the green circle in Figure \ref{fig:qtmt_map}. Its partition path can be expressed as three \gls{qt} splits (\gls{qt} depth 3), followed by a \gls{hbt} split and two \gls{ns} decisions.

\section{CNN-based Prediction of Partition Path}\label{4}

% \textcolor{red}{Paragraph 25 offering a transition between the representation and its prediction by CNN}
Predicting the optimal partition is equivalent to predicting the optimal partition path. In \gls{vtm}, the size of \gls{ctu} is set to 128x128 by default, consequently yielding QTdepthMap and MTsplitMap dimensions of 16x16 and 32x32, respectively. The representation of partition path can be predicted by a multi-branch \gls{cnn}, where one branch infers the QTdepthMap of regression values with dimension 16x16x1, while the other three branches produce the MTsplitMap. Each element of MTsplitMap is classified into one of five classes, corresponding to five split types, resulting in three MT outputs with dimensions of 32x32x5. We have handled the classification of \gls{mt} splits as an image segmentation problem based on 4x4 sub-blocks. Accordingly, we adopted the classical U-Net structure \cite{ronneberger2015u} to design our \gls{cnn} model to address this segmentation-like task.

% \textcolor{red}{Paragraph 26 giving an outline of this section}
In this section, the U-Net structure is briefly introduced. Then we present the structure of the proposed \gls{cnn} in Section \ref{4.2}. Afterwards, we list its input features and explain the reasons for choosing them in Section \ref{4.3}.

\subsection{U-Net}\label{4.1}
% \textcolor{red}{Paragraph 27 introducing briefly U-net}
The U-Net structure is derived from \gls{fcn}\cite{long2015fully}. It consists of an encoder part which is composed of a sequence of convolutional layers plus maxpooling layers. Then this part is followed by a decoder part in which the maxpooling layers are replaced by upsampling layers. In addition, skip connections concatenate feature maps from the encoder and decoder with the same dimension. The U-Net and its variations have been widely applied to image segmentation tasks.

\subsection{MS-MVF-CNN Structure} \label{4.2}

\begin{figure*}
    \centering
    \includegraphics[width=0.88\linewidth]{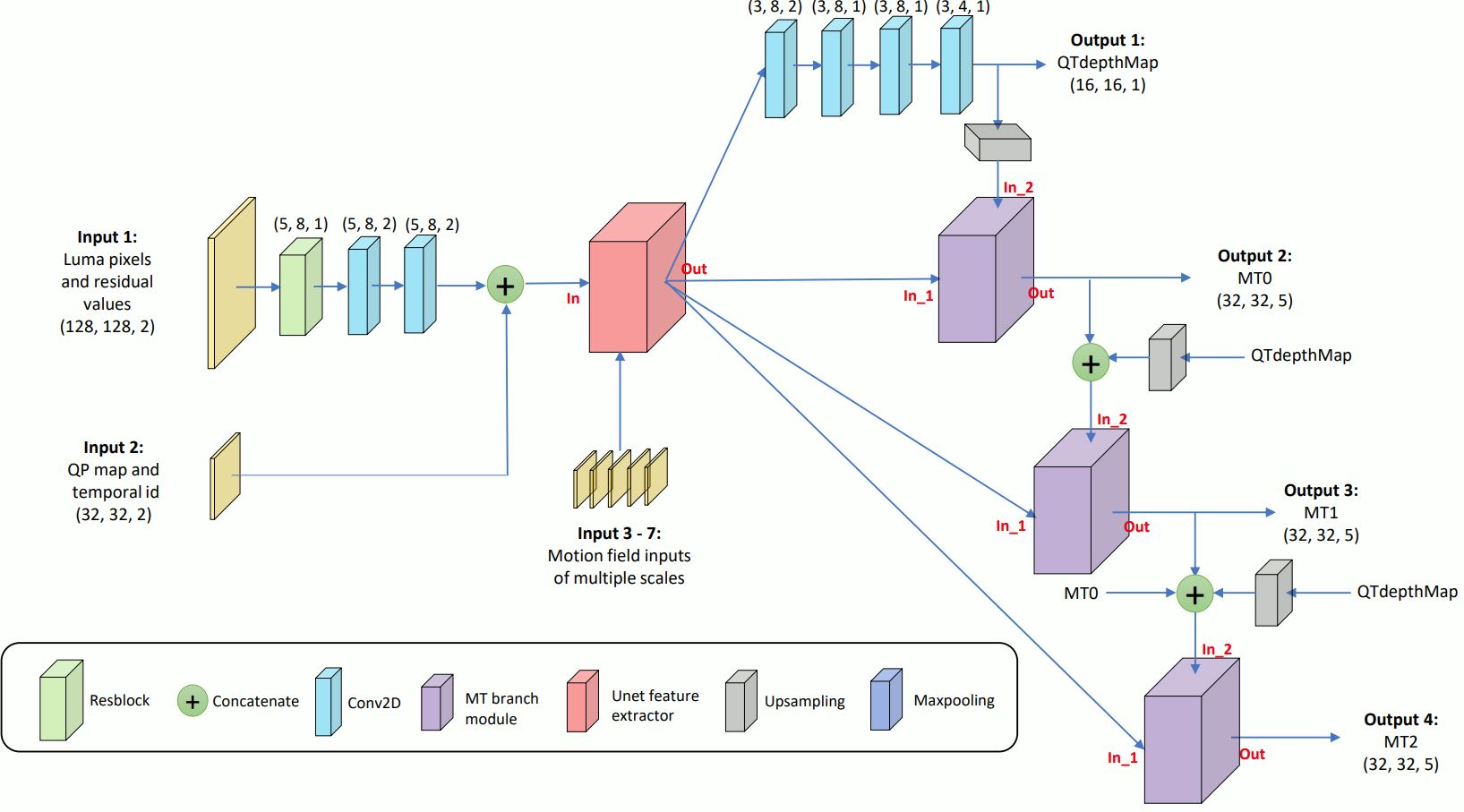}
    \caption{Multi-Scale Motion Vector Field CNN. The vector above Resblock and Conv2D represents (kernel size, number of filters, stride).}
    \label{fig:CNN_main}
\end{figure*}

\begin{figure*}
    \centering
    \includegraphics[width=0.88\linewidth]{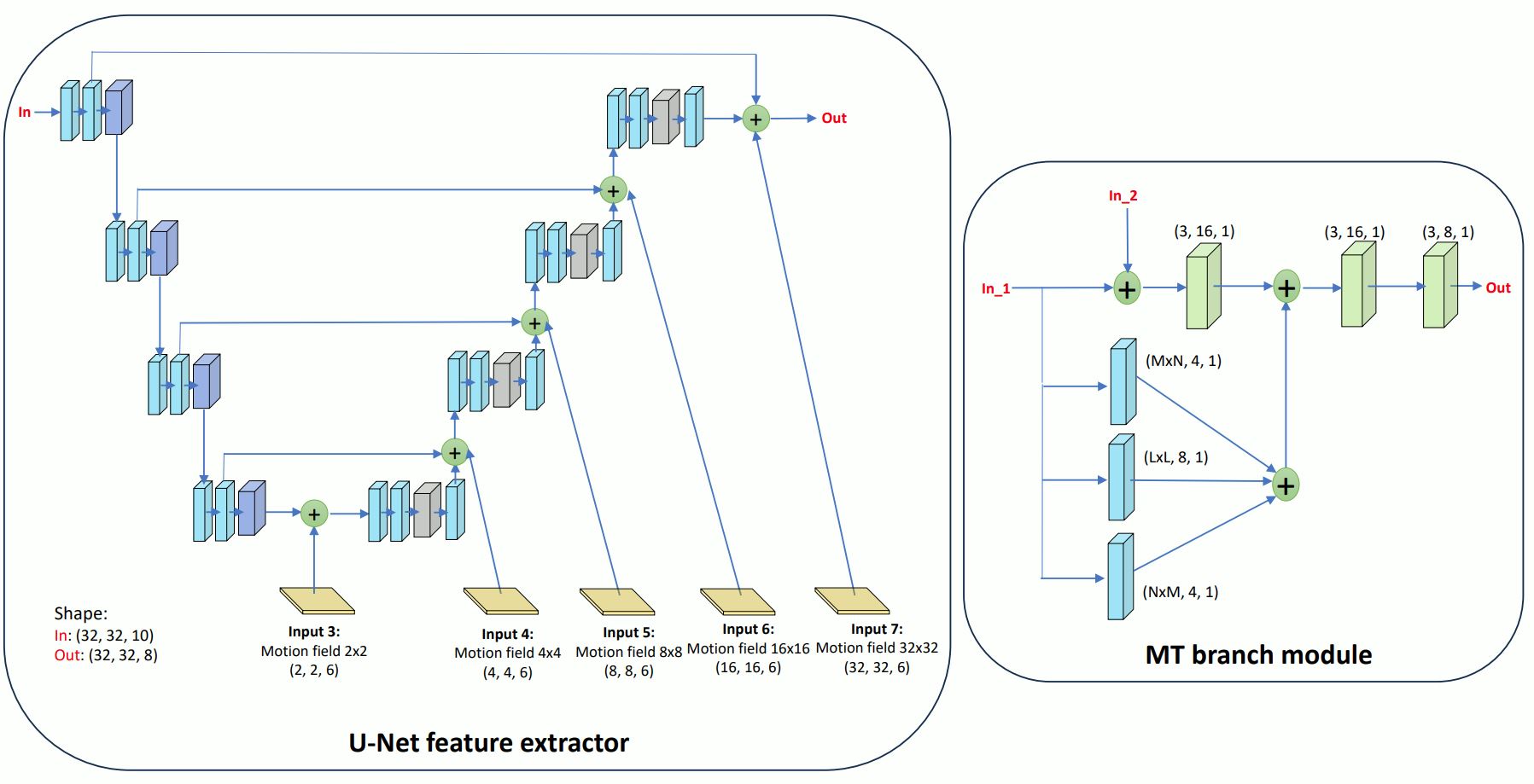}
    \caption{U-Net feature extractor and MT branch module}
    \label{fig:CNN_module}
\end{figure*}

% \textcolor{red}{Paragraph 28 describing the U-Net feature extractor part}
The \gls{cnn} structure proposed in this paper, named \gls{msmvfcnn}, is depicted in Figure \ref{fig:CNN_main}.
The proposed \gls{cnn} has 7 inputs and 4 outputs. After two convolutional layers with stride, the tensor of Input 1 is downsampled to dimension 32x32x8 and then concatenated with the Input 2. The merged input is then fed to the U-Net feature extractor demonstrated in Figure \ref{fig:CNN_module}. Regarding the design of this module, we are referring to the classical structure of U-Net depicted in \cite{ronneberger2015u}. Specifically, we concatenate the upsampled feature map in the decoder part of U-Net, the feature map copied from the encoder part with the motion vector field of the same scale. At the decoding part, the feature map is gradually expanded and merged with normalized motion field of 2x2x6, 4x4x6, 8x8x6, 16x16x6 and 32x32x6. As a result, the U-Net feature extractor outputs a feature map of dimension 32x32x8, combining pixels features with motion estimation features.

% \textcolor{red}{Paragraph 29 describing the CNN part for predicting QT and MT maps}
Since the split at each level depends on previous splits, we employ a hierarchical multi-branch prediction mechanism. QTdepthMap is predicted after shrinking the features extracted from U-Net by four convolutional layers. For \gls{mt} branches, we designed the MT branch module presented in Figure \ref{fig:CNN_module}. Two inputs of this module are the extracted features of U-Net and outputs from previous partition levels. We utilize the asymmetric kernel structure to process the extracted features. This structure is originally proposed by \cite{chen2020learned} in \gls{hevc} to pay attention to near-horizontal and near-vertical textures for predicting split decision of intra coding by \gls{cnn}. We adopt this structure to exploit the horizontal and vertical information contained in \gls{msmvf}. The MT branch module contains branches of kernel size MxN, LxL, and NxM. The values of (M, N, L) are set as (5, 7, 9) for branch MT0, (3, 5, 7) for branch MT1, and (1, 3, 3) for branch MT2. On deeper \gls{mt} levels, splits are made on smaller \gls{cu}s. Thus, smaller kernel sizes are applied to extract finer features. After the asymmetric kernels, the feature map is then concatenated with outputs from previous levels. In the end, the merged feature maps are given to two residual blocks \cite{he2016deep} before yielding classification results of \gls{mt} branches. No activation is applied to the fully connected output layer of the \gls{qt} depth branch. The output layer of the \gls{mt} branch is with softmax activation.

\subsection{Input Features} \label{4.3}

% \textcolor{red}{Paragraph 30 giving a short transition}
This network structure takes three different types of input. The involved inputs are presented below:

\subsubsection{Original and Residual CTU}\label{4.3.1}

% \textcolor{red}{Paragraph 31 presenting the first input}
In Figure \ref{fig:CNN_main}, Input 1, with dimensions of 128x128x2, is created by merging the original \gls{ctu} with the residual \gls{ctu}. The original luma pixels carry the texture details of the \gls{ctu}, while the residual \gls{ctu} is generated through motion compensation of the original \gls{ctu} based on the nearest frame.

% \textcolor{red}{Paragraph 32 explaining the reason for using the first input}
Several studies \cite{pan2021cnn, yeo2021cnn, wang2018fast} have adopted a method in which both the original \gls{ctu} values and the residual of \gls{ctu} are fed to a \gls{cnn}. Combining the original and residual values as input allows \gls{cnn} to assess the similarity between current \gls{ctu} and reference \gls{ctu}. This combined input offers features that reflect the temporal correlation between frames which is a crucial factor in inter partition prediction.

\subsubsection{QP and Temporal ID}\label{4.3.2}

% \textcolor{red}{Paragraph 33 presenting the second input}
The Input 2, as illustrated in Figure \ref{fig:CNN_main}, has dimensions of 32×32×2, consisting of two separate 32x32 matrices. These matrices are assigned specific values: one holds the \gls{qp} value, while the other contains the temporal identifier. This temporal identifier in \gls{vvc}, similar to its usage in \gls{hevc}, signifies a picture's position within a hierarchical temporal prediction structure, controlling temporal scalability \cite{wang2021high}.

% \textcolor{red}{Paragraph 34 explaining the reason for using the second input}
We specifically utilize the \gls{qp} value and temporal identifier as input features since inter partitioning depends on them. In essence, a higher temporal layer identifier or a lower \gls{qp} value tends to result in finer partitions, as outlined in \cite{yeo2021cnn}. Instead of developing separate models for each parameter instance, our approach focuses on training a model with adaptability to varying values of \gls{qp} and temporal identifier.

\subsubsection{Multi-Scale Motion Vector Field}\label{4.3.3}

% \textcolor{red}{Paragraph 35 presenting the multiscale mv field input}
In this paper, we have introduced a \gls{cnn} model based on a novel input feature called \gls{msmvf}. Our \gls{msmvf} at five scales is presented as Input 3-7 in Figure \ref{fig:CNN_module}. To compute \gls{msmvf}, we divide the 128x128 \gls{ctu} into multiple scale sub-blocks ranging from 4x4 pixels to 64x64 pixels, and perform motion estimation on these sub-blocks. Each motion vector of sub-block comprises a vertical and horizontal motion value, along with the associated \gls{sad} cost value as the third element. By concatenating elements pointing to reference frame of L0 with those of L1, each sub-block corresponds to 6 elements in the motion vector field. For example, the motion vector field input for 8x8-pixel scale has dimensions of 16x16x6.

% \textcolor{red}{Paragraph 36 presenting mv related features in the state of the art}
A significant challenge in inter partition prediction is the large motion search space, which spans up to 6 regions of 384x384 pixels across different reference frames in the \gls{ragop32} configuration. State-of-the-art methods typically employ motion fields or pixels from reference frames as input features for machine learning models. Notably, in \cite{amestoy2019tunable} and \cite{pan2021cnn}, a crucial feature used is the motion field, which comprises motion vectors calculated for each 4x4 sub-block referring to the nearest frame. As mentioned in \cite{amestoy2019tunable}, this motion field is strongly correlated with the optimal partition. In a different approach, Tissier \textit{et al.} in \cite{tissier2022machine} opt to utilize two reference \gls{ctu}s in the nearest frames.

% \textcolor{red}{Paragraph 37 presenting the motivation for adopting it as input}
The choice of using \gls{msmvf} as the \gls{cnn} input, instead of motion fields and reference pixels, is based on the following reasons. First, the \gls{msmvf} contains crucial motion information for the current \gls{ctu}, which is essential for both inter prediction and inter partitioning. This information can be interpreted more effectively by the \gls{cnn} model compared to using reference pixels as \gls{cnn} input. Second, the multi-scale nature of \gls{msmvf} aligns well with the multi-level structure of U-Net and can leverage this structure effectively. Essentially, \gls{msmvf} represents motion features at various resolutions, allowing for the combination with features extracted from \gls{ctu} pixels at the same resolution scale.

\begin{figure}[h]
    \centering
    \begin{tikzpicture}
\begin{axis}[
	xlabel={BD-rate loss \scriptsize{($\%$)}},
	ylabel={Percentage of encoding time saving \scriptsize{($\%$)}},
	legend pos=south east,
	legend style={nodes={scale=0.7, transform shape}},
	width=9cm,
	y label style={at={(axis description cs:0.1,.5)},anchor=south},
	grid=major
]
\addplot coordinates {
(2.25,  37.0)
(4.15,	45.9)
(6.68,	52.1)
(11.57,	57)
};

\addplot[green,mark=square] coordinates {
(1.8,   40.4)
(3.31,	47.0)
(5.69,	51.2)
(9.58,	57.7)

};

% \node [black] at (axis cs:0.5, 16) {\scriptsize{Th:0.4}};
% \node [black] at (axis cs:1.2,	26) {\scriptsize{Th:0.1}};
% \node [black] at (axis cs:1.6,  31) {\scriptsize{Th:0}};

\legend{PIX-CNN \scriptsize{(VTM10)}, MVF-CNN \scriptsize{(VTM10)},\scriptsize{(VTM10)}}
\end{axis}
\end{tikzpicture}
    \caption{Comparison of performances between PIX-CNN and MVF-CNN.
    }
    \label{fig:compare_input}
\end{figure}
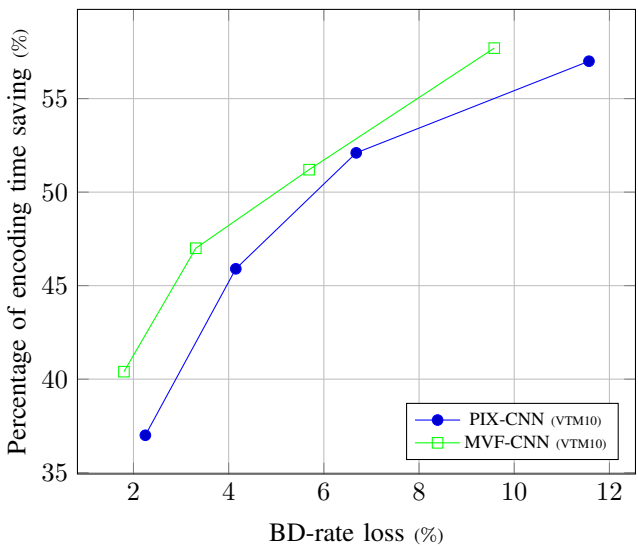

% \textcolor{red}{Paragraph 38 demonstrating the advantage of such input over pixel input by experiment}
To demonstrate the effectiveness of our \gls{msmvf} input, we conducted an experiment involving the training of two \gls{cnn} models. The only distinction between these models is their input: the first model, PIX-CNN, takes the pixels of two reference \gls{ctu}s as input, while the second model, MVF-CNN, utilizes our proposed \gls{msmvf} as input. Both models share the same architecture as in Figure \ref{fig:CNN_main}. The training dataset comprises 250k samples randomly selected from the \gls{ragop32} encoding of 200 sequences with a resolution of 540p from \cite{ma2021bvi}. Performance evaluations in Figure \ref{fig:compare_input} are based on Class C sequences of \gls{ctc}. The results consistently show that MVF-CNN outperforms PIX-CNN at all four data points,  which justifies the advantages of using the \gls{msmvf} input over pixel input.

% \textcolor{red}{Paragraph 39 offering a complexity analysis for generating this CNN input}
Based on our evaluation conducted on the first 64 frames of all \gls{ctc} sequences using the \gls{ragop32} configuration, the computation of \gls{msmvf} for each \gls{ctu} consumes, on average, a mere 0.52\% of the encoding time in VTM10. Importantly, the generation of \gls{msmvf} introduces only minimal encoding overhead, making it a task that can be readily preprocessed or parallelized.

\section{Proposed CNN-based Acceleration Method} \label{5}

% \textcolor{red}{Paragraph 40 providing a transition between CNN prediction and the method for using it}
After the prediction by our trained \gls{cnn} model, we obtain one QTdepthMap and three MTsplitMaps per \gls{ctu}. The predicted QTdepthMap is composed of floating-point values. The predicted MTsplitMaps comprise probabilities of five split types for each 4x4 sub-block within the \gls{ctu}. In this section, we elucidate the post-processing of the \gls{cnn} prediction, with the aim of achieving a wide range of acceleration-loss trade-off.

\begin{algorithm}
	\caption{MT splits early skipping}
      \label{alg:mt_skipping}
	 \textbf{Input:} \\\text{QTdepthMap; MTsplitMap; Thm; QTdepth\textsubscript{cur}}, \text{ CU; Size\textsubscript{CU}; Pos\textsubscript{CU}}\\
     \textbf{Output:} \\\text{SkipMT: Boolean to decide whether to skip MT split}\\\text{types or not.}\\
     \text{CandSplit: Candidate list of splits for RDO check}
	\begin{algorithmic}[1]
        \State Compute the average QTdepth\textsubscript{pred} based on Size\textsubscript{CU}, Pos\textsubscript{CU} and QTdepthMap
        
        \If{round(QTdepth\textsubscript{pred}) $>$ QTdepth\textsubscript{cur} \textbf{and} \newline
         QT is possible for current CU}
        \State SkipMT = True
        \State CandSplit = \{NS, QT\}
        \Else 
        \State SkipMT = False
        \State CandSplit = \{NS\}
         \For{sp in \{BTH, BTV, TTH, TTV\}}
        \State Compute average Proba\textsubscript{sp} based on Size\textsubscript{CU}, Pos\textsubscript{CU} and MTsplitMap
         \If{Proba\textsubscript{sp} $>$ Thm}
         \State CandSplit append split sp
         \EndIf
         \EndFor    
		\EndIf
	\end{algorithmic} 
\end{algorithm} 

% \textcolor{red}{Paragraph 41 explaining why we build a cand list per level}
Decision errors at low partitioning depth can result in large loss of \gls{bdrate}. Based on the predictions of our \gls{cnn}, selecting the best single partition path, equivalent to choosing the best split at each \gls{mt} level, will not be optimal or scalable. Our approach involves generating candidate lists at each level, which means that multiple partition paths are chosen for the \gls{rdo} test. This approach of creating candidate lists at various levels is designed to achieve satisfying trade-off between acceleration and coding loss while assuring the scalabilty of method.

\begin{figure}[h]
\centering
    \includegraphics[width=0.6\linewidth]{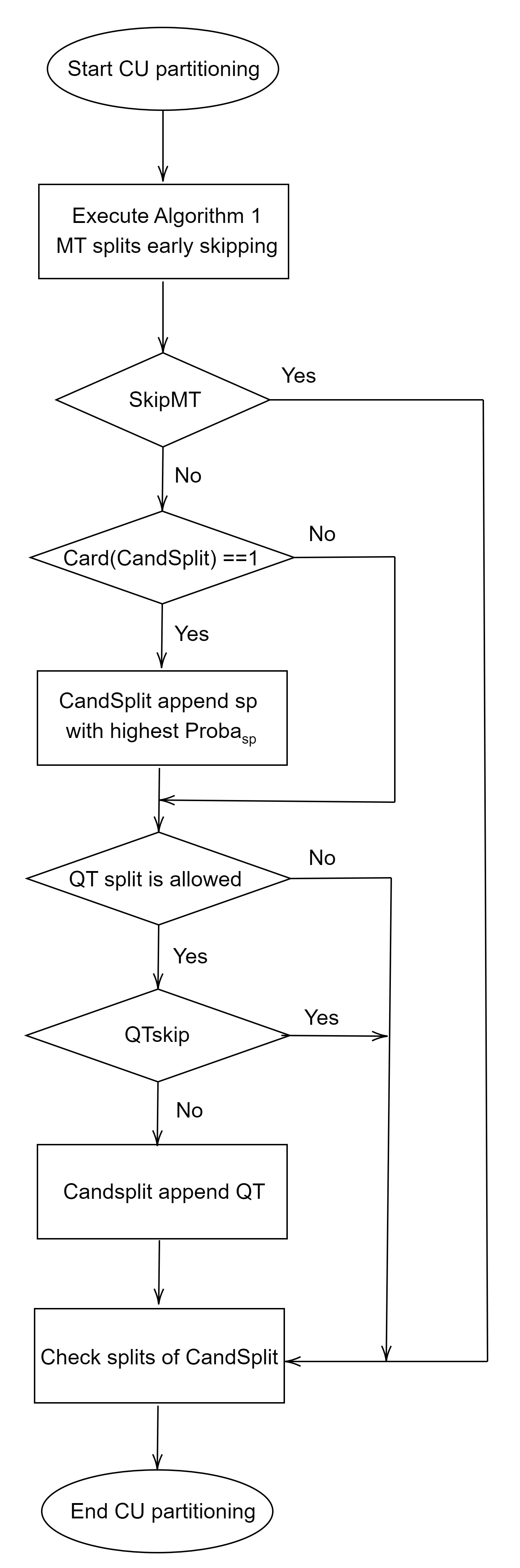}
    \caption{Flowchart of acceleration algorithm}
    \label{fig:flowchart}
\end{figure}

% \textcolor{red}{Paragraph 42 introducing the parameters for adjusting the trade off}
The acceleration algorithm is precisely described in Algorithm \ref{alg:mt_skipping} and Figure \ref{fig:flowchart}. We introduce two parameters \emph{Thm} and \emph{QTskip} to regulate the acceleration-loss trade-off. Specifically, \emph{Thm} is the threshold for the split probability. \emph{QTskip} represents whether we should accelerate \gls{rdo} of \gls{qt} splits or not. Increasing the \emph{Thm} value and setting \emph{QTskip} to true will lead to greater acceleration at the cost of increased coding loss.

% \textcolor{red}{Paragraph 43 presenting the algo executed at the CU level}
Regarding the algorithm applied at the \gls{cu} level, Algorithm. \ref{alg:mt_skipping} is first executed. This algorithm produces two outputs: the \emph{SkipMT} variable and the \emph{CandSplit} list, both of which are subsequently utilized in the flowchart in Figure \ref{fig:flowchart}. To start with, the mean \emph{QTdepth\textsubscript{pred}} of current \gls{cu} is calculated based on the corresponding area in \gls{qtdepthmap}. If the rounded \emph{QTdepth\textsubscript{pred}} is larger than the QT depth of the \gls{cu} and \gls{qt} split is feasible, the current \gls{cu} should be split by \gls{qt}. Consequently, all \gls{mt} splits are excluded from the \emph{CandSplit} list and \emph{SkipMT} is set to true. Otherwise, the mean probability of each available split is computed on the corresponding MTsplitMap in a similar way to that of the QTdepth. Then \emph{CandSplit} is filled by splits with \emph{Proba\textsubscript{sp}} larger than the threshold \emph{Thm}. In this case, the value assigned to \emph{SkipMT} is False.

% \textcolor{red}{Paragraph 44 presenting the execution of flowchart following the Algo}
In the flowchart of Figure \ref{fig:flowchart}, if the \emph{SkipMT} is true after the execution of Algorithm \ref{alg:mt_skipping}, we directly check the \emph{CandSplit}. In this scenario, the encoder conducts \gls{rdo} of \gls{cu} and splits \gls{cu} with \gls{qt} because \emph{CandSplit} contains only \gls{ns} and \gls{qt}. If \emph{SkipMT} is false, then we will verify if \gls{ns} is the only choice in \emph{CandList}. If this is the case, we will add the \gls{mt} split with the highest probability to the list. Next, if \gls{qt} split is not allowed for \gls{cu} due to \gls{cu} shape or shortcuts, we go directly to the check of \emph{CandSplit}. If the \gls{qt} split is feasible, we refer to \emph{QTskip} to determine whether to add \gls{qt} to the \emph{CandList} or not. Setting the \emph{QTskip} to true signifies that we will always check \gls{qt} if possible. This is for rectifying the potential error of predicting a \emph{QTdepth\textsubscript{pred}} value smaller than the actual ground truth value. However, it comes at the expense of sacrificing some acceleration. Finally, we execute \gls{rdo} on \gls{cu} and partition it by split types in the \emph{CandSplit} list. The partition search then repeats for the next \gls{cu}, and the algorithm described above is applied anew.

% \textcolor{red}{Paragraph 45 concluding this section}
Our inter partitioning acceleration method is designed on top of the partitioning algorithm of \gls{vtm} which performs a nearly exhaustive search on possible partition paths of a \gls{ctu}, except that it incorporates a handful of handcrafted conditional shortcuts as mentioned in Section \ref{2}. Therefore, this work can be considered as a \gls{cnn}-based shortcut to reduce the search space of partition paths.

\section{Training of MS-MVF-CNN} \label{6}

% \textcolor{red}{Paragraph 46 giving an outline of this section}
To effectively train our \gls{cnn} model, we have designed a hybrid loss function and created a large-scale dataset named MVF-Inter\footref{mvf_inter}. First, we will explain how this loss function is determined in Section \ref{6.1}. Then Section \ref{6.2} describes training details and the generation of dataset.

\subsection{Loss Function}\label{6.1}

% \textcolor{red}{Paragraph 47 introducing the loss function}
The outputs of \gls{msmvfcnn} contain one regression output as well as three classification outputs. Therefore, a hybrid loss function is developed in our case. We choose the category cross-entropy for classification loss and mean square error for regression loss as follows: 

\begin{equation}
L = a \frac{1}{n_q} \sum_{i=1}^{n_q}(d_i-\hat{d_i})^2 - (1-a)(\sum_{b=1}^{n_b}\sum_{i=1}^{n_m}\sum_{s=1}^{n_s}w_{b,s}y_{b,i,s}\log(\hat{y}_{b,i,s}))  
\label{eq:loss_1}
\end{equation}

% \textcolor{red}{Paragraph 48 introducing the variables in the loss function}
Here, we have \emph{n\textsubscript{q}} = 256, \emph{n\textsubscript{b}} = 3, \emph{n\textsubscript{m}} = 1024 and \emph{n\textsubscript{s}} = 5, representing the number of elements in QTdepthMap, the number of MT branches, the number of elements in MTsplitMap, and the number of split types, respectively. In this equation, \emph{d\textsubscript{i}} denotes the ground-truth QT depth value, while \emph{\^{d}\textsubscript{i}} represents the predicted QT depth value. Additionally, \emph{\^{y}\textsubscript{b,i,s}} is used to denote the predicted probability of split type \emph{s} for the \emph{i}-th element of the \gls{mt} decision map at the \emph{b}-th \gls{mt} branch. Similarly, \emph{y\textsubscript{b,i,s}} signifies the ground-truth label for the same case. Notably, we introduce a parameter \emph{a}, which falls within the range [0, 1], in Equation \ref{eq:loss_1} to fine-tune the relative weights of the regression loss and classification loss.

\begin{figure}[h]
    \includegraphics[width=1\linewidth]{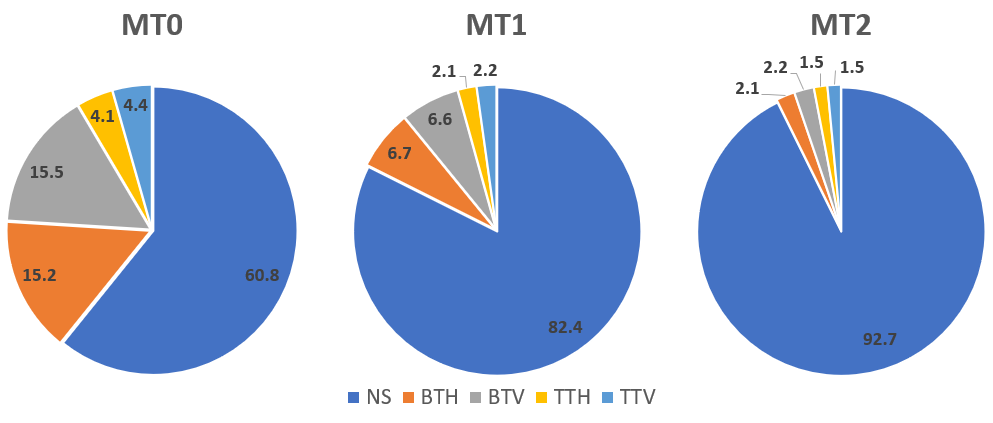}
    \caption{Distribution of split types for MT0, MT1, MT2}
    \label{fig:dis_mt}
\end{figure}

% \textcolor{red}{Paragraph 49 presenting the class weights in the loss function}
The split types are distributed unbalancedly at different \gls{mt} depths as illustrated in Figure \ref{fig:dis_mt}. To counteract this imbalance, we introduce class weights for split type \emph{s} on \gls{mt} branch \emph{b}, denoted as \emph{w\textsubscript{b,s}}. The definition of these weights is as follows:

\begin{equation}
w_{b,s} = \frac{\lambda_{s}p_{b,s=ns}}{p_{b,s}}   
\label{eq:loss_weight}
\end{equation}

% \textcolor{red}{Paragraph 50 explaining how these class weights are obtained}
where \emph{p\textsubscript{b,s}} represents the percentage of split type \emph{s} within \gls{mt} branch \emph{b}. For each branch \emph{b}, $\frac{p\textsubscript{b,s=ns}}{p\textsubscript{b,s}}$ can be interpreted as the inverse percentage of the split type \emph{s} normalized by the inverse percentage of the \gls{ns} split. In \cite{huang2021block}, a series of tests were performed to evaluate the coding gain and increase of complexity associated with the \gls{bt} and \gls{tt} splits individually as demonstrated in Table \ref{tab:bt_tt_test}.

\begin{table}[!h]
    \centering
    \caption{Settings of split type in VTM9 under RA \cite{huang2021block}}
    \begin{tabular}{c|c|c|c|c}
    \toprule  \hline
         \multirow{5}{*} & BT split & TT split & BD-rate & Encoding Time\\ \cline{1-5}
         Anchor Setting & X & X & - & - \\\cline{1-5}
         Setting 1 & \checkmark & X & -8.26\% & 337\% \\\cline{1-5} Setting 2 & X & \checkmark & -10.22\% & 732\% \\\cline{1-5}

    \bottomrule
    \end{tabular}
    \label{tab:bt_tt_test}
\end{table}

% \textcolor{red}{Paragraph 51 explaining why the bt class deserves a higher class weight}
When comparing Setting 1 and Setting 2 to the anchor configuration, it's observed that Setting 1 and Setting 2 exhibit similar BD-rate gains, but the encoding time in Setting 2 is twice that of Setting 1. These tests suggest that \gls{bt} split performs better in terms of the trade-off between complexity and coding gain compared to \gls{tt} split. 
Thus, placing greater importance on the prediction of the \gls{bt} split can result in a better acceleration-loss trade-off. To achieve this, the ratio between the proportion of \gls{ns} and proportion of split \emph{s} is computed for \gls{mt} branch \emph{b}. The class weight \emph{w\textsubscript{b,s}} in Equation \ref{eq:loss_weight} is formulated as the product of this ratio and $\lambda\textsubscript{s}$ which is another weight added to prioritize the split type \emph{s}.

% \textcolor{red}{Paragraph 52 showing the final loss function after the tuning}
After fine-tuning the model, we find that the best performance is achieved with a value of 0.8 for \emph{a} and $\lambda\textsubscript{s}$ set to 2 for \gls{bt} splits and 1 for \gls{tt} splits and \gls{ns}.

\subsection{Dataset Generation and Training Details} \label{6.2}

% \textcolor{red}{Paragraph 53 outlining the value for generating this dataset}
Constructing a large scale inter partition dataset is more challenging than that of intra partition because the former needs to encode a substantial number of video sequences, while the latter could be done by encoding images. To the best of our knowledge, there exists no prior work focused on developing an inter partition dataset.

% \textcolor{red}{Paragraph 54 presenting how the dataset is generated}
Our MVF-Inter\footref{mvf_inter} dataset involved the encoding of 800 sequences from \cite{ma2021bvi} and an additional 28 sequences of 600 frames in 720p resolution extracted from \cite{wang2019youtube}. Sequences of \cite{ma2021bvi} cover resolutions of 240p, 540p, 1080p, and 4k, with 200 videos of 64 frames for each resolution. We have encoded all these videos with the \gls{vtm}10\cite{vtm10} encoder in the \gls{ragop32} configuration with \gls{qp} 22, 27, 32, and 37. We randomly selected a total of 820k \gls{ctu} partition samples, equally distributed per resolution and \gls{qp}, with 120k samples reserved as a validation set.

% \textcolor{red}{Paragraph 55 presenting the form of data contained in each sample}
Each sample of our dataset contains the following components of each \gls{ctu}: pixel values, residual values, motion vector fields at five scales, \gls{qp} value, temporal ID value, QTdepthMap with depths ranging from 0 to 4, and MTsplitMaps for MT0, MT1, and MT2. MTsplitMap labels are encoded as \gls{vtt} (0), \gls{vbt} (1), \gls{ns} (2), \gls{hbt} (3), and \gls{htt} (4).

% \textcolor{red}{Paragraph 56 offering some training details}
In terms of training details, we employed the Adam optimizer~\cite{kingma2014adam} to train the model. The initial learning rate was set to 1e$^{-3}$ and was exponentially decreased by 3\% every 5 epochs. The batch size set for training is 400.

\section{Experimental Results and Analyses} \label{7}

% \textcolor{red}{Paragraph 57 giving outline of this section}
In this section, we present the results of our experiments and provide an in-depth analysis of the results. To begin, in Section \ref{7.1}, we assess the precision of the prediction of our \gls{cnn} model. Subsequently, comparisons with the \gls{rf} and \gls{cnn} based approaches are made in Section \ref{7.2}. Finally, the complexity analysis of our framework is carried out in Section \ref{7.3}.

\subsection{Prediction Accuracy Evaluation} \label{7.1}
\begin{figure}[h]
    \includegraphics[width=1\linewidth]{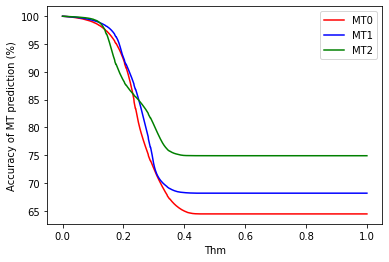}
    \caption{Curves of accuracy and \emph{Thm} for MT0, MT1, MT2}
    \label{fig:mt_accuracy}
\end{figure}

% \textcolor{red}{Paragraph 58 presenting the two decisions to evaluate}
At the \gls{cu} level, our algorithm can be broken down into two decisions: the decision of \emph{SkipMT} and the decision of \emph{CandSplit} list. To evaluate the precision of decisions based on our model's output, we have performed the encoding where both the ground truth partitioning and the \gls{cnn} output were collected. The analysis is done on the first 64 frames of all \gls{ctc} sequences excluding class D with \gls{qp} 22, 27, 32, 37. The accuracy of these decisions presented in Table \ref{tab:skipMT} and Figure \ref{fig:mt_accuracy} are calculated by averaging four \gls{qp}s and various test sequences.

% \textcolor{red}{Paragraph 59 explaining how the first decision is evaluated}
There is no need to make a \emph{SkipMT} decision on \gls{qt} depth 4 since the partitioning is forced to proceed to \gls{mt} splits with the maximum of \gls{qt} depth reached. The accuracy of \emph{SkipMT} decision is independently measured on \gls{qt} depth from 0 to 3. If the current \gls{cu} requires further splitting of \gls{qt} and \emph{SkipMT} is equal to False, then this decision of \emph{SkipMT} is classified as False Negative (FN). The proportion of True Positives (TP), FN, True Negatives (TN), False Positives (FP) and their corresponding Precision (Prec) and Recall (Rec) are shown in Table \ref{tab:skipMT}. Precision, recall, and F1 score are calculated as follows:

\begin{equation}
Precision\textsubscript{QTdepth} = \frac{TP\textsubscript{QTdepth}}{TP\textsubscript{QTdepth} + FP\textsubscript{QTdepth}}   
\label{eq:precision}
\end{equation}

\begin{equation}
Recall\textsubscript{QTdepth} = \frac{TP\textsubscript{QTdepth}}{TP\textsubscript{QTdepth} + FN\textsubscript{QTdepth}}   
\label{eq:recall}
\end{equation}

\begin{equation}
F1score\textsubscript{QTdepth} = 2 \frac{Precision\textsubscript{QTdepth} Recall\textsubscript{QTdepth} }{Precision\textsubscript{QTdepth} + Recall\textsubscript{QTdepth}}   
\label{eq:F1}
\end{equation}

% \textcolor{red}{Paragraph 60 showing observations from the evaluation of the first decision}
Generally, our model exhibits strong performance at \gls{qt} depths ranging from 0 to 2, as depicted in Table \ref{tab:skipMT}. Both precision and F1 score decrease as \gls{qt} depth increases. At \gls{qt} depth 3, the precision and F1 score drop to 25\% and 40\%, respectively, suggesting that the \emph{SkipMT} decision at this level is less reliable. These observations could be explained by two reasons:

% \textcolor{red}{Paragraph 61 offering first explanation for the observation}
First of all, the scale of decision-making diminishes as the QT depth increases. More explicitly, the \emph{SkipMT} decision at \gls{qt} depth 0 is made at the \gls{ctu} scale by computing the mean of 256 values from the QTdepthMap. Nevertheless, the decision at \gls{qt} depth 3 relies only on 4 values from the QTdepthMap within the 16x16 \gls{cu}. Consequently, decisions at smaller scales are less resilient to incorrectly predicted QTdepthMap values, resulting in lower overall accuracy at higher QT depths.

% \textcolor{red}{Paragraph 62 offering second explanation for the observation}
Secondly, decisions at higher \gls{qt} depths are noticeably more imbalanced than those at lower \gls{qt} depths. Positive cases of ground truth at \gls{qt} depth 3 represent only 0.02\% , while the proportion of positive cases is 49.65\% at QT depth 0. In conclusion, the model is trained in such a way that it tends to make negative \emph{SkipMT} decision at larger \gls{qt} depths. This explains the decline in precision as the \gls{qt} depth increases.

\begin{table}[!h]
    \centering
    \caption{Table of confusion for \emph{SkipMt} (Unit: \%)}
    \begin{tabular}{c|c|c|c|c|c|c|c}
    \toprule  \hline
         \multirow{5}{*} & TP & FN & TN & FP & Prec & Rec & F1score\\ \cline{1-8}
         QT depth 0  & 41.84 & 7.81 & 45.83 & 4.52 & 90.3 & 84.3 & 87.2\\\cline{1-8} QT depth 1 & 19.53 & 0.58 & 72.57 & 7.32 & 72.7 & 97.1 & 83.1\\\cline{1-8}
         QT depth 2 & 2.69 & 0.08 & 94.67 & 2.57 & 51.1 & 97.1 & 67.0\\\cline{1-8}
         QT depth 3 & 0.02 & 0 & 99.92 & 0.06 & 25.0 & 100.0 & 40.0\\\cline{1-8}

    \bottomrule
    \end{tabular}
    \label{tab:skipMT}
\end{table}

% \textcolor{red}{Paragraph 63 showing observations and its explanation from the evaluation of the second decision}
In Figure \ref{fig:mt_accuracy}, the accuracy of the \emph{CandSplit} list decision is determined by whether the list contains the ground truth split at the \gls{mt} level. We calculate and draw separate accuracy curves for MT0, MT1 and MT2 separately by varying the threshold \emph{Thm}. As \emph{Thm} increases, the size of the \emph{CandSplit} list decreases, leading to decreasing precision. Once \emph{Thm} reaches a certain value, the accuracy stabilizes because \emph{CandList} is constant, containing only the \gls{mt} split type with the highest probability and \gls{ns}. It's worth noting  that the minimum accuracy of the \gls{mt} increases with the \gls{mt} depth. This is mainly due to the fact that \gls{ns} is more frequent at larger \gls{mt} depths, as illustrated in the pie chart in Figure \ref{fig:dis_mt}. Since our \emph{CandSplit} list consistently includes \gls{ns}, the accuracy tends to be relatively higher at larger \gls{mt} depths.

% \textcolor{red}{Paragraph 64 concluding the evaluation of prediction accuracy}
In general, our model achieves a satisfactory F1 score for QT depths 0, 1 and 2 regarding the \emph{SkipMT} decision. As for the  \emph{CandSplit} list decision, our algorithm maintains an accuracy exceeding 65\% while adjusting the value of \emph{Thm} at various \gls{mt} levels. These performance evaluations justifies the high accuracy of the decisions made by our method during the partition search process in \gls{vvc}.

\subsection{Comparison with the State of the Art} \label{7.2}

\begin{figure}[h]
    \centering
    \begin{tikzpicture}
\begin{axis}[
	xlabel={BD-rate loss \scriptsize{($\%$)}},
	ylabel={TS \scriptsize{($\%$)}},
	legend pos=south east,
	legend style={nodes={scale=0.7, transform shape}},
	width=9cm,
	y label style={at={(axis description cs:0.1,.5)},anchor=south},
	grid=major
]
\addplot[red,mark=asterisk] coordinates {
(0.435,  16.5) % (False, 0)
(1.231,	35) % (True, 0)
(1.34,	40.6) %(True, 0.125)
(1.98,	49.9) %(True, 0.175)
(3.22,	56.5) %(True, 0.225)
(4.59,	60.2) %(True, 0.26)
};

\addplot[black,mark=triangle] coordinates {
(1.40,  38) % (False, 0)
(3.24,	51) % (True, 0)
(4.65,	56) % (True, 0.125)
};

\addplot[green,mark=diamond] coordinates {
(1.16,  18.5) % (medium)
(2.824,	50.1) % (fast)
};

% \addplot[black,mark=diamond] coordinates {
% (1.11,  31.8) % (medium)
% (2.33,	43.4) % (fast)
% (3.12,	54.3) % (fast)
% };

\node [black] at (axis cs:0.25, 18) {\scriptsize{(F, 0)}};
\node [black] at (axis cs:0.9,	35) {\scriptsize{(T, 0)}};
\node [black] at (axis cs:1.0,  43) {\scriptsize{(T, 0.125)}};
\node [black] at (axis cs:1.5,	51) {\scriptsize{(T, 0.175)}};
\node [black] at (axis cs:2.8,  56) {\scriptsize{(T, 0.225)}};
\node [black] at (axis cs:4,  62) {\scriptsize{(T, 0.26)}};

\node [black] at (axis cs:1.8,  37) {\scriptsize{(T, 0.125)}};
\node [black] at (axis cs:3.7,  50) {\scriptsize{(T, 0.225)}};
\node [black] at (axis cs:4.6,  53) {\scriptsize{(T, 0.26)}};

\node [black] at (axis cs:1.6, 18) {\scriptsize{Medium}};
\node [black] at (axis cs:2.5, 50) {\scriptsize{Fast}};

\legend{Ours \scriptsize{(VTM10)}, 
Ours \scriptsize{(VTM21)}, 
Reproduction of \cite{kulupana2021fast} \scriptsize{(VTM10)},\scriptsize{(VTM10)}}
\end{axis}
\end{tikzpicture}
    \caption{Comparison of performances between proposed method and reproduction of \cite{kulupana2021fast}.
    }
    \label{fig:compare_rf}
\end{figure}
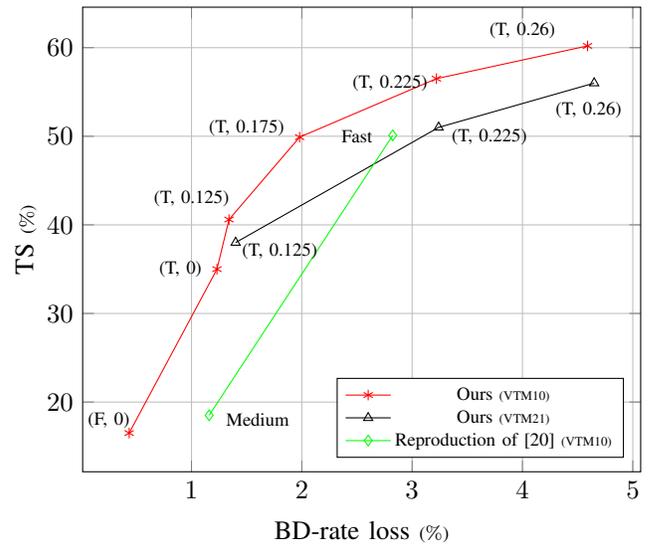

% \textcolor{red}{Paragraph 65 explaining the configuration for experiments}
The proposed method has been implemented in the \gls{vtm}10.0 encoder using the Frugally deep library \cite{frugallydeep2018} for CPU-based inference in real time. To showcase the effectiveness of our method in the latest version of \gls{vtm}, we conducted experiments using \gls{vtm}21, as represented by the black curve in Figure \ref{fig:compare_rf}. Encodings of \gls{ctc} sequences are performed on a Linux machine with Intel Xeon E5-2697 v4 in a single-threaded manner. These experiments were conducted on the first 64 frames of \gls{ctc} sequences with the \gls{ragop32} configuration on four \gls{qp}s values of 22, 27, 32, 37.

% \textcolor{red}{Paragraph 66 presenting the metrics for the experiment}
Two metrics were used to assess the performance: BD-rate \cite{bjontegaard2001calculation} and Time Saving (TS). The formula for computing TS is provided in Equation \ref{eq:ts}. Here, T\textsubscript{Test} denotes the encoding time of the proposed method, while T\textsubscript{VTM} represents the encoding time of the original \gls{vtm}10 under the same conditions. The average BD-rate loss and \gls{ts} are computed as the arithmetic mean and geometric mean, respectively, on four \gls{qp}s values over \gls{ctc} sequences as defined in \cite{jvet_ctc}. In addition, sequences of class D are excluded when computing the overall average performance.

\begin{equation}
TS = \frac{1}{4} \sum_{q \in \{22,27,32,37 \}} \frac{T_{VTM}(q) - T\textsubscript{Test}(q)}{T_{VTM}(q)}
\label{eq:ts}
\end{equation}

% \textcolor{red}{Paragraph 67 explaining why we cant directly compare our result with rf base methods}
The acceleration performances obtained from the state-of-the-art \gls{rf}-based methods could not be directly compared with our performance. There are two main reasons for this. First of all, the results of \cite{amestoy2019tunable} and \cite{kulupana2021fast} are based on \gls{vtm}5.0 and \gls{vtm}8.0, respectively. The differences of encoder complexity among various \gls{vtm} versions are not negligible as highlighted in \cite{vtmcomplexity}, which makes it less valid to directly compare our performances with theirs. Secondly, the training dataset was generated from a subset of \gls{ctc} sequences, and the results were not obtained from the entire \gls{ctc}. This approach results in possible overfitting and reduces the credibility of their results. As a result, comparing our results obtained on the entire \gls{ctc} with their results is not fair. 

% \textcolor{red}{Paragraph 68 explaining how we have reproduced the rf methods}
\cite{kulupana2021fast} is an extended and specialized work for \gls{vvc} based on \cite{amestoy2019tunable}. We have reproduced the result of \cite{kulupana2021fast} in \gls{vtm}10 to perform an unbiased comparison between our method  and \gls{rf}-based method in \cite{kulupana2021fast}. First of all, we created a non-\gls{ctc} dataset for training. Table \ref{tab:dataset_seq} presents details on the composition of sequences for the dataset. For the 720p resolution, sequences are selected from \cite{wang2019youtube} and sequences for other resolutions are from \cite{ma2021bvi}. In the end, we generated a large dataset with 3.7e$^{7}$ samples for the training of 17 Hor/Ver classifiers as well as 2.5e$^{6}$ samples for the training of 4 QT/MTT classifiers. After generating the dataset, we trained, pruned and integrated the \gls{rf} classifiers in \gls{vtm}10.0. This was done in a manner consistent with the original article, including the implementation of the early termination rule for \gls{tt}
\footnote{The code and dataset of reproduction is available at \url{https://github.com/Simon123123/vtm10\_fast\_dt\_inter\_partition\_pcs2021.git
}}.

\begin{table}[!h]
    \centering
    \caption{Breakdown of sequences used to train \gls{rf}s of \cite{kulupana2021fast}}
    \begin{tabular}{c|c|c|c|c|c}
    \toprule
         &  \multicolumn{5}{c}{Resolution} \\ \hline
         \multirow{2}{*}{\shortstack{Number of\\videos}} & 240p & 480p & 720p & 1080p & 4k \\ \cline{2-6}
         & 50 & 13 & 10 & 10 & 5\\\cline{2-6}

    \bottomrule
    \end{tabular}
    \label{tab:dataset_seq}
\end{table}

\begin{table*}[t]
\centering
\caption{\label{tab:Result}Performance of the proposed method in comparison with reference CNN-based methods (Unit: \%)}        
\begin{tabular}{c|c|cc|cc|cc|cc|cc|cc}
\hline 
\hline
\multirow{2}{*}{Class} &    \multirow{2}{*}{Sequence}      & \multicolumn{2}{c|}{Pan \cite{pan2021cnn} \scriptsize{(VTM6)}} & \multicolumn{2}{c|}{Tissier\cite{tissier2022machine} \scriptsize{(C2)}} & \multicolumn{2}{c|}{Tissier\cite{tissier2022machine} \scriptsize{(C3)}} & \multicolumn{2}{c|}{Ours \scriptsize{(T, 0, VTM6)}} & \multicolumn{2}{c|}{Ours \scriptsize{(T, 0.125, VTM10)}} & \multicolumn{2}{c}{Ours \scriptsize{(T, 0.2, VTM10)}} \\
                       &  & BD-rate            & TS   & BD-rate            & TS          & BD-rate              & TS              & BD-rate            & TS            & BD-rate               & TS & BD-rate & TS \\
\hline 
\parbox[t]{2mm}{\multirow{7}{*}{\rotatebox[origin=c]{90}{A \scriptsize{(4k)} }}} 	& {\footnotesize Tango2} 			& 4.03 			& 38.56 			& - 			& -   & - 	& -		& 1.35 			& 32.3 			& 1.84 			& 43.7  & 3.08 & 57.5 \\
																								& {\footnotesize FoodMarket4} 		& 1.74 			& 46.12 			& - 			& - 		& - 	& -			& 0.75 			& 29.4 			& 0.85 			& 55.1 & 1.13 & 53.6\\
																								& {\footnotesize Campfire} 			& 3.17 			& 38.23 			& - 			& - 			& - 	& -		& 1.49 			& 40.7			& 1.83 			& 48.5& 3.22 & 63.2 \\
																								& {\footnotesize CatRobot1} 		& 6.45 			& 36.84 			& - 			& - 		& - 	& -		& 1.31 			& 36.5 			& 1.45 			& 42.6 & 2.67 & 57.6 \\
																								& {\footnotesize DaylightRoad2} 	& 5.63 			& 35.47 			& -			& - 	
                        & - 	& -	 & 1.57			& 39.4			& 2.00 			& 45.6 & 3.94 & 57.9\\
																								& {\footnotesize ParkRunning3} 		& 2.10 			& 26.45 			& -			& - 		& - 	& -			& 0.99 			& 42.6 			& 0.98 			& 45.9 & 1.93 & 59.8 \\
\cline{2-14}	                                                                                                                                                                                            
																										   & {\textbf{Average}} 	& \textbf{3.85} & \textbf{36.46} 	& \textbf{1.84} & \textbf{47.7} 	& \textbf{3.06} & \textbf{59.7} 	& \textbf{1.25} & \textbf{37.0} & \textbf{1.49} & \textbf{45.3} &\textbf{2.66} & \textbf{58.4} \\
\hline 	                                                                                                                                                                                                    
	                                                                                                                                                                                                        
\parbox[t]{2mm}{\multirow{6}{*}{\rotatebox[origin=c]{90}{B \scriptsize{(1080p)} }}} 	& {\footnotesize MarketPlace} 		& 4.33 			& 33.64 			& - 			& - 			& - 	& -	  & 0.99 			& 37.6 			& 1.48 			& 46.3  & 2.78 & 57.7\\
																								& {\footnotesize RitualDance} 		& 3.55 			& 34.17 			& - 			& - 
                        & - 	& -	 & 1.75			& 39.9 			& 1.91 			& 49.4 & 3.91 & 61.8\\
																								& {\footnotesize Cactus} 			& 5.72 			& 29.36 			& - 			& - 
                        & - 	& -	 & 1.05 			& 37.8 			& 1.30 			& 44.8 & 2.45 & 58.3 \\
																								& {\footnotesize BasketballDrive} 	& 3.30 			& 37.28 			& - 			& -
                        & - 	& -	 & 1.34 			& 39.6 			& 1.95 			& 49.7 & 3.65 & 63.4 \\
																								& {\footnotesize BQTerrace} 		& 1.90 			& 20.21 			& - 			& - 	
                       & - 	& -	  & 0.99 			&32.6 			& 1.18			& 39.8 & 2.23 & 52.2 \\
\cline{2-14}                                                                                                                                                                                                
																								& \textbf{Average} 	& \textbf{3.76} & \textbf{30.27} 	& \textbf{2.21} & \textbf{46.5} 	& \textbf{3.09} & \textbf{58.2} 	& \textbf{1.22} & \textbf{37.5} & \textbf{1.56} & \textbf{46.1} &\textbf{3.00} &\textbf{58.9} \\
\hline                                                                                                                                                                                                      
\parbox[t]{2mm}{\multirow{5}{*}{\rotatebox[origin=c]{90}{C \scriptsize{(480p)} }}} 	& {\footnotesize BasketballDrill} 	& 2.29 			& 29.23 			& - 			& - 			& - 	& -		& 1.04 			& 26.9 			& 1.08 			& 30.3 & 2.60 & 39.7 \\
																								& {\footnotesize BQMall} 			& 2.69 			& 27.48 			& - 			& - 
                        & - 	& -				& 1.20 			& 29.1 			& 1.18 			& 32.2 & 2.71 & 39.7\\
																								& {\footnotesize PartyScene} 		& 2.22 			& 20.80 			& - 		& - 	
                        & - 	& -	  & 0.78 			& 31.5 			& 0.86 			& 33.3 & 2.25 & 43.3\\
																								& {\footnotesize RaceHorses} 		& 3.02 			& 26.39 			& - 			& - 	
                        & - 	& -	  & 0.96 			& 32.8			& 1.09 			& 34.6 & 2.94 & 45.6\\
\cline{2-14}                                                                                                                                                                                                
																								& \textbf{Average} 					& \textbf{2.56} & \textbf{25.77} 	& \textbf{3.20} & \textbf{43.1} 	& \textbf{3.79} & \textbf{53.8} 	& \textbf{0.99} & \textbf{30.1} &\textbf{1.05} &\textbf{32.6} & \textbf{2.63} & \textbf{42.1}\\
\hline                                                                                                                                                                                      																	\parbox[t]{2mm}{\multirow{5}{*}{\rotatebox[origin=c]{90}{D \scriptsize{(240p)} }}} 	& {\footnotesize BasketballPass} 	& 1.85 			& 26.97 			& - 			& - 				& - 	& -	 & 0.76 			& 19.0 			& 0.85 			& 22.2 & 1.72 & 25.1 \\
																								& {\footnotesize BQSquare} 			& 1.61 			& 14.86 			& - 			& - 	
                        & - 	& -	 & 0.50 			& 17.6			& 0.54 			& 19.6 & 1.38 & 22.5 \\
																								& {\footnotesize BlowingBubbles} 	& 3.03 			& 22.15 			& - 			& - 	
                        & - 	& -	 & 0.33 			& 17.2 			& 0.45 			& 18.9  & 1.12 & 23.4 \\
																								& {\footnotesize RaceHorses} 		& 2.92 			& 24.20 			& - 			& - 	
                        & - 	& -	   & 1.04			& 22.8 			& 0.85 			& 24.4 & 2.11 & 31.2 \\
\cline{2-14}                                                                                                                                                                                                
																								& \textbf{Average} 					& \textbf{2.35} & \textbf{21.53} 	& \textbf{3.02} & \textbf{36.8} 	& \textbf{3.26} & \textbf{45.2} 	& \textbf{0.66} & \textbf{19.2} & \textbf{0.67} & \textbf{21.0} & \textbf{1.58} & \textbf{25.6} \\		\hline	                                                                                                                           
 \parbox[t]{2mm}{\multirow{4}{*}{\rotatebox[origin=c]{90}{E \scriptsize{(720p)} }}}  	& {\footnotesize FourPeople} 		& 2.31 			& 33.77 			& -			& -			
 & - 	& -	  & 0.95 			& 29.5 			& 0.90 			& 34.3 & 1.65 & 41.2 \\
																								& {\footnotesize Johnny} 			& 3.53 			& 35.22 			& -			& -	
                        & - 	& -	   & 0.93 			& 22.1 			& 1.13 			& 27.6 & 2.01 & 32.8 \\
																								& {\footnotesize KristenAndSara} 	& 2.58 			& 36.50 			& -			& -	
                        & - 	& -	   & 1.00 			& 24.3 			& 1.11 			& 30.3 & 1.73 & 36.4 \\
\cline{2-14}                                                                                                                                                                                                
																								& \textbf{Average} 					& \textbf{2.81} & \textbf{35.15} 	& \textbf{1.45}	& \textbf{38.7}		& \textbf{2.2} & \textbf{49.6} 	& \textbf{0.96} & \textbf{25.4} &\textbf{1.04} & \textbf{30.8} & \textbf{1.79} &\textbf{36.9}\\
\hline

\hline

          		\multicolumn{2}{c|}{\textbf{Total average}} & \textbf{3.18} & \textbf{30.63} 	& \textbf{2.33} & \textbf{43.4} 	& \textbf{3.12} & \textbf{54.3} 	& \textbf{1.14} & \textbf{33.8} & \textbf{1.34} & \textbf{40.6} &\textbf{2.60} &\textbf{52.2}\\

\hline

 \hline
\end{tabular}
\label{tab:cnn_compare}
\end{table*}

% \textcolor{red}{Paragraph 69 comparing our methods with rf methods after reproduction}
We reproduce the result of the medium and fast speed preset of \cite{kulupana2021fast} in \gls{vtm}10. It should be noted that the maximum \gls{mt} depth is limited to 2 for the fast preset. We plot the curve of BD-rate loss and \gls{ts} of our method by gradually adjusting the threshold \emph{Thm} and \emph{QTskip} to build six settings. The curves obtained are shown in Figure \ref{fig:compare_rf}. For example, the label (T, 0.125) signifies that in this particular setting, \emph{QTskip} and \emph{Thm} are assigned the values True and 0.125, respectively. Our method can achieve scalable acceleration varying from 16.5\% to 60.2\% with BD-rate loss ranging from 0.44\% to 4.59\%. Comparing with the fast preset, the setting (T, 0.175) produces the same acceleration with a 0.84\% lower BD-rate loss. Similarly, the setting (T, 0) reaches the same BD-rate loss while providing a 17\% higher speed-up compared to the medium preset. In summary, our method generally outperforms the state-of-the-art \gls{rf}-based method. It is worth mentioning that the results in \gls{vtm}21 are obtained by implementing our \gls{cnn} model, which was originally trained on \gls{vtm}10. Consequently, it is expected to exhibit reduced performance compared to the results in \gls{vtm}10. Nonetheless, our method remains applicable and effective in the latest version of \gls{vtm}.

% \textcolor{red}{Paragraph 70 comparing our methods with CNN methods}
Regarding \gls{cnn}-based approaches, we compared our method with \cite{pan2021cnn} and \cite{tissier2022machine} in Table \ref{tab:cnn_compare}. The \gls{vtm} version of \cite{pan2021cnn} is \gls{vtm}6. Thus we reimplement our method and integrate our model trained on \gls{vtm}10 into \gls{vtm}6 for a fair comparison within the same context. In Table \ref{tab:cnn_compare}, the reimplementation in \gls{vtm}6 labeled as (T, 0, \gls{vtm}6) reaches a slightly larger acceleration with only one-third of BD-rate loss compared to \cite{pan2021cnn}. For \cite{tissier2022machine}, their \gls{vtm} version is the same as ours, allowing for direct comparisons. Encoding with \emph{Thm} = 0.125 yields a 40. 6\% reduction in the encoding time, which is similar to the acceleration achieved by the C2 configuration in \cite{tissier2022machine}, but with only half of its BD-rate loss. Furthermore, our method with \emph{Thm} set to 0.2 outperformed their C3 configuration, achieving a 0.52\% lower BD-rate loss at the same level of acceleration. In conclusion, our method consistently outperforms all state-of-the-art methods.

% \textcolor{red}{Paragraph 71 observation from experiments and explanation for it}
It is important to note that the level of acceleration can vary depending on different sequence classes (e.g. resolution), which is consistent with other \gls{cnn}-based methods. As discussed in \cite{huang2021block}, \gls{ctu}s that exceed the picture boundary are called partial \gls{ctu}s. These partial \gls{ctu}s require a different partition search scheme compared to regular \gls{ctu}s. Consequently, the encoding of partial \gls{ctu}s are not accelerated since the \gls{cnn}-based approaches are not applicable to them. Generally, the proportion of the frame region occupied by partial \gls{ctu}s is larger for lower resolutions, resulting in less acceleration when fast partitioning approaches are used on smaller resolutions. This could partially explain the limited acceleration observed in class D which was excluded from the overall performance calculation. More specifically, our method tends to perform better on higher resolutions (e.g. class A and class B) while achieving less acceleration than state-of-the-art methods on lower resolutions (e.g. class C, class D and class E). Investigating and improving this aspect could be a focus of future work.

\subsection{Complexity Analysis} \label{7.3}

% \textcolor{red}{Paragraph 72 explaining the necessity for designing light methods}
Machine learning-based fast partitioning methods may not be suitable for alternative implementations of the same codec. For example, VVenc~\cite{vvenc} is a fast implementation of \gls{vvc}. In the All Intra configuration, \gls{vtm}10.0 is reported to be 27 times more complex compared to VVenc with fast preset, as mentioned in \cite{vvencai}. The overall complexity of the \gls{cnn}-based method presented in \cite{Feng2023cnn} accounts for only 2.34\% of the encoding time of the \gls{vtm}10 encoder. However, when this method is implemented in VVenc without any adjustments, its overhead increases to about 67\% of the encoding time with the fast preset, which means that this method is not directly applicable to VVenc. Consequently, it is crucial to develop a lightweight method to ensure its applicability across different implementations. Furthermore, lightweight methods do not require parallel execution, enhancing the cost-effectiveness of such solutions.

\begin{table}[!h]
    \centering
    \caption{Overhead of our method (Unit: \%)}
    \begin{tabular}{c|c|c|c|c|c|c}
    \toprule  \hline
         \multirow{3}{*} & 240p & 480p & 720p & 1080p & 4k & Average \\ \cline{1-7}
         CNN  & 0.23 & 0.37 & 0.99 & 0.90 & 0.84  & 0.60\\\cline{1-7} Preprocess & 0.24 & 0.41 & 1.15 & 0.81 & 0.86 & 0.62\\\cline{1-7}

    \bottomrule
    \end{tabular}
    \label{tab:complexity}
\end{table}

% \textcolor{red}{Paragraph 73 analysing the complexity of our method in terms of encoding time}
As a result,  we conducted a complexity analysis of our method to compare it with the state of the art. The overhead of a machine learning-based method typically consists of three components: preprocessing time, inference time, and postprocessing time. The post-processing of our method is integrated into the \gls{vvc} partitioning process and introduces minimal overhead to the encoding process. However, preprocessing is necessary to compute the \gls{msmvf} as model input. Table \ref{tab:complexity} provides the complexity of the preprocessing and the inference of \gls{cnn} related to the encoding of the anchor \gls{vtm}10. The last column corresponds to the geometric average of complexity for sequences from class A to E (including class D). Based on experimental results, the \gls{cnn} inference time on a CPU accounts for only 0.60\% of the total encoding time. Our approach consumes only 1.21\% of the total encoding time, underscoring its lightweight nature.

% \textcolor{red}{Paragraph 74 analysing the complexity of our method and comparing with existing methods in terms of flop}
Another important metric for evaluating the complexity of the model is its \gls{flops}. Our model has a \gls{flops} value of 1.12e$^{6}$. In comparison, the \gls{flops} of the model in \cite{li2021deepqtmt} is approximately 1.1e$^{9}$\cite{wu2022hg}. \cite{wu2022hg} employs a pruned ResNet-18 as the backbone with 9e$^{7}$ \gls{flops}, and \cite{tissier2022machine} utilizes the pretrained MobileNetV2 with 3.14e$^{8}$ \gls{flops}. Our model is hundreds of times lighter than these methods. The lightweight nature of our proposed approach facilitates its adaptation to faster encoders.

\section{Conclusion} \label{8}

% \textcolor{red}{Paragraph 75 giving final conclusion}
In this study, we propose a machine learning-based method to accelerate \gls{vvc} inter partitioning. Our method leverages a novel representation of the \gls{qtmt} partition structure based on partition path, consisting of QTdepthMap and MTsplitMaps. Our work is structured as follows. Firstly, we have built a large scale inter partition dataset. Secondly, a novel Unet-based model that takes \gls{msmvf} as input is trained to predict the partition paths of \gls{ctu}. Thirdly, we develop a scalable acceleration algorithm based on thresholds to utilize the output of the model. Finally, we speed up the \gls{vtm}10 encoder under \gls{ragop32} configuration by 16.5\%$\sim$60.2\% with \gls{bdrate} loss of 0.44\%$\sim$4.59\%. This performance surpasses state-of-the-art methods in terms of coding efficiency and complexity trade-off. Notably, our method is among the most lightweight methods in the field, making it possible to adapt our approach to faster codecs.

% \textcolor{red}{Paragraph 76 listing furture works}
For future work, we intend to investigate how video resolution influences partitioning acceleration, aiming to boost the speed-up of our method on lower resolutions. Furthermore, there is still acceleration potential lying in the selection of inter coding modes at the \gls{cu} level, as discussed in \cite{liu2022icip}. An extension of our approach could be the incorporation of fast inter coding mode selection algorithm into our method to further accelerate the inter coding process.

\bibliographystyle{unsrt}
\bibliography{ref}

\begin{thebibliography}{10}

\bibitem{cisco}
Cisco.
\newblock {Cisco Annual Internet Report (2018–2023) White Paper}, 2020.

\bibitem{bross2021overview}
B.~Bross, Y.~Wang, Y.~Ye, S.~Liu, J.~Chen, G.J. Sullivan, and J.~Ohm.
\newblock Overview of the versatile video coding (vvc) standard and its
  applications.
\newblock {\em IEEE Transactions on Circuits and Systems for Video Technology},
  31(10):3736--3764, 2021.

\bibitem{wang2016adaptive}
Z.~Wang, J.~Zhang, N.~Zhang, and S.~Ma.
\newblock Adaptive motion vector resolution scheme for enhanced video coding.
\newblock In {\em 2016 Data Compression Conference (DCC)}, pages 101--110,
  2016.

\bibitem{li2017efficient}
L.~Li, H.~Li, D.~Liu, Z.~Li, H.~Yang, S.~Lin, H.~Chen, and F.~Wu.
\newblock An efficient four-parameter affine motion model for video coding.
\newblock {\em IEEE Transactions on Circuits and Systems for Video Technology},
  28(8):1934--1948, 2018.

\bibitem{alshin2010bi}
A.~Alshin, E.~Alshina, and T.~Lee.
\newblock Bi-directional optical flow for improving motion compensation.
\newblock In {\em 28th Picture Coding Symposium}, pages 422--425, 2010.

\bibitem{huang2021block}
Y.W. Huang, J.~An, H.~Huang, X.~Li, S.T. Hsiang, K.~Zhang, H.~Gao, J.~Ma, and
  O.~Chubach.
\newblock Block partitioning structure in the vvc standard.
\newblock {\em IEEE Transactions on Circuits and Systems for Video Technology},
  31(10):3818--3833, 2021.

\bibitem{complexity}
A.~Tissier, A.~Mercat, T.~Amestoy, W.~Hamidouche, J.~Vanne, and D.~Menard.
\newblock Complexity reduction opportunities in the future vvc intra encoder.
\newblock In {\em 2019 IEEE 21st International Workshop on Multimedia Signal
  Processing (MMSP)}, pages 1--6. IEEE, 2019.

\bibitem{fan2020fast}
Y.~Fan, J.~Chen, H.~Sun, J.~Katto, and M’E. Jing.
\newblock A fast qtmt partition decision strategy for vvc intra prediction.
\newblock {\em IEEE Access}, 8:107900--107911, 2020.

\bibitem{cui2020gradient}
J.~Cui, T.~Zhang, C.~Gu, X.~Zhang, and S.~Ma.
\newblock Gradient-based early termination of cu partition in vvc intra coding.
\newblock In {\em 2020 Data Compression Conference (DCC)}, pages 103--112,
  2020.

\bibitem{chen2019vcip}
J.~Chen, H.~Sun, J.~Katto, X.~Z., and Y.~Fan.
\newblock Fast qtmt partition decision algorithm in vvc intra coding based on
  variance and gradient.
\newblock In {\em 2019 IEEE Visual Communications and Image Processing (VCIP)},
  pages 1--4, 2019.

\bibitem{lei2019look}
M.~Lei, F.~Luo, X.~Zhang, S.~Wang, and S.~Ma.
\newblock Look-ahead prediction based coding unit size pruning for vvc intra
  coding.
\newblock In {\em 2019 IEEE International Conference on Image Processing
  (ICIP)}, pages 4120--4124, 2019.

\bibitem{saldanha2020fast}
M.~Saldanha, G.~Sanchez, C.~Marcon, and L.~Agostini.
\newblock Fast partitioning decision scheme for versatile video coding
  intra-frame prediction.
\newblock In {\em 2020 IEEE International Symposium on Circuits and Systems
  (ISCAS)}, pages 1--5, 2020.

\bibitem{fu2019icme}
T.~Fu, H.~Zhang, F.~Mu, and H.~Chen.
\newblock Fast cu partitioning algorithm for h.266/vvc intra-frame coding.
\newblock In {\em 2019 IEEE International Conference on Multimedia and Expo
  (ICME)}, pages 55--60, 2019.

\bibitem{galpin2019cnn}
F.~Galpin, F.~Racap{\'e}, S.~Jaiswal, P.~Bordes, F.~Le~L{\'e}annec, and
  E.~Fran{\c{c}}ois.
\newblock Cnn-based driving of block partitioning for intra slices encoding.
\newblock In {\em 2019 Data Compression Conference (DCC)}, pages 162--171.
  IEEE, 2019.

\bibitem{tissier2022machine}
A~Tissier, W~Hamidouche, J~Vanne, and D~Menard.
\newblock Machine learning based efficient qt-mtt partitioning for vvc inter
  coding.
\newblock In {\em 2022 IEEE International Conference on Image Processing
  (ICIP)}, pages 1401--1405. IEEE, 2022.

\bibitem{wu2022hg}
S.~Wu, J.~Shi, and Z.~Chen.
\newblock Hg-fcn: Hierarchical grid fully convolutional network for fast vvc
  intra coding.
\newblock {\em IEEE Transactions on Circuits and Systems for Video Technology},
  32(8):5638--5649, 2022.

\bibitem{Feng2023cnn}
A.~Feng, K.~Liu, D.~Liu, L.~Li, and F.~Wu.
\newblock Partition map prediction for fast block partitioning in vvc
  intra-frame coding.
\newblock {\em IEEE Transactions on Image Processing}, 32:2237--2251, 2023.

\bibitem{saldanha2021configurable}
M.~Saldanha, G.~Sanchez, C.~Marcon, and L.~Agostini.
\newblock Configurable fast block partitioning for vvc intra coding using light
  gradient boosting machine.
\newblock {\em IEEE Transactions on Circuits and Systems for Video Technology},
  32(6):3947--3960, 2021.

\bibitem{amestoy2019tunable}
T.~Amestoy, A.~Mercat, W.~Hamidouche, D.~Menard, and C.~Bergeron.
\newblock Tunable vvc frame partitioning based on lightweight machine learning.
\newblock {\em IEEE Transactions on Image Processing}, 29:1313--1328, 2020.

\bibitem{kulupana2021fast}
G.~Kulupana, V.P. Kumar~M, and S.~Blasi.
\newblock Fast versatile video coding using specialised decision trees.
\newblock In {\em 2021 Picture Coding Symposium (PCS)}, pages 1--5, 2021.

\bibitem{pan2021cnn}
Z.~Pan, P.~Zhang, B.~Peng, N.~Ling, and J.~Lei.
\newblock A cnn-based fast inter coding method for vvc.
\newblock {\em IEEE Signal Processing Letters}, 28:1260--1264, 2021.

\bibitem{yeo2021cnn}
W.~Yeo and B.G. Kim.
\newblock {CNN-based Fast Split Mode Decision Algorithm for Versatile Video
  Coding (VVC) Inter Prediction}.
\newblock {\em Journal of Multimedia Information System}, 8(3):147--158, 2021.

\bibitem{liu2022light}
Y.~Liu, M.~Abdoli, T.~Guionnet, C.~Guillemot, and A.~Roumy.
\newblock Light-weight cnn-based vvc inter partitioning acceleration.
\newblock In {\em 2022 IEEE 14th Image, Video, and Multidimensional Signal
  Processing Workshop (IVMSP)}, pages 1--5. IEEE, 2022.

\bibitem{tissier2023machine}
A.~Tissier, W.~Hamidouche, SBD. Mdalsi, J.~Vanne, F.~Galpin, and D.~Menard.
\newblock Machine learning based efficient qt-mtt partitioning scheme for vvc
  intra encoders.
\newblock {\em IEEE Transactions on Circuits and Systems for Video Technology},
  2023.

\bibitem{wieckowski2019fast}
A~Wieckowski, J~Ma, H~Schwarz, D~Marpe, and T~Wiegand.
\newblock Fast partitioning decision strategies for the upcoming versatile
  video coding (vvc) standard.
\newblock In {\em 2019 IEEE International Conference on Image Processing
  (ICIP)}, pages 4130--4134. IEEE, 2019.

\bibitem{rdcost}
G.J. Sullivan and T.~Wiegand.
\newblock Rate-distortion optimization for video compression.
\newblock {\em IEEE Signal Processing Magazine}, 15(6):74--90, 1998.

\bibitem{ronneberger2015u}
O~Ronneberger, P~Fischer, and T~Brox.
\newblock U-net: Convolutional networks for biomedical image segmentation.
\newblock In {\em Medical Image Computing and Computer-Assisted
  Intervention--MICCAI 2015: 18th International Conference, Munich, Germany,
  October 5-9, 2015, Proceedings, Part III 18}, pages 234--241. Springer, 2015.

\bibitem{long2015fully}
J~Long, E~Shelhamer, and T~Darrell.
\newblock Fully convolutional networks for semantic segmentation.
\newblock In {\em Proceedings of the IEEE conference on computer vision and
  pattern recognition}, pages 3431--3440, 2015.

\bibitem{chen2020learned}
Z.~Chen, J.~Shi, and W.~Li.
\newblock Learned fast hevc intra coding.
\newblock {\em IEEE Transactions on Image Processing}, 29:5431--5446, 2020.

\bibitem{he2016deep}
K.~He, X.~Zhang, S.~Ren, and J.~Sun.
\newblock Deep residual learning for image recognition.
\newblock In {\em Proceedings of the IEEE conference on computer vision and
  pattern recognition}, pages 770--778, 2016.

\bibitem{wang2018fast}
Z.~Wang, S.~Wang, X.~Zhang, S.~Wang, and S.~Ma.
\newblock Fast qtbt partitioning decision for interframe coding with
  convolution neural network.
\newblock In {\em 2018 25th IEEE International Conference on Image Processing
  (ICIP)}, pages 2550--2554, 2018.

\bibitem{wang2021high}
Y.~Wang, R.~Skupin, M.~Hannuksela, S.~Deshpande, V.~Drugeon, R.~Sj{\"o}berg,
  B.~Choi, V.~Seregin, Y.~Sanchez, J.~Boyce, et~al.
\newblock The high-level syntax of the versatile video coding (vvc) standard.
\newblock {\em IEEE Transactions on Circuits and Systems for Video Technology},
  31(10):3779--3800, 2021.

\bibitem{ma2021bvi}
D.~Ma, F.~Zhang, and D.R. Bull.
\newblock Bvi-dvc: A training database for deep video compression.
\newblock {\em IEEE Transactions on Multimedia}, 24:3847--3858, 2021.

\bibitem{wang2019youtube}
Y.~Wang, S.~Inguva, and B.~Adsumilli.
\newblock Youtube ugc dataset for video compression research.
\newblock In {\em 2019 IEEE 21st International Workshop on Multimedia Signal
  Processing (MMSP)}, pages 1--5. IEEE, 2019.

\bibitem{vtm10}
Y.~Ye J.~Chen and S.~Kim.
\newblock Algorithm description for versatile video coding and test model 10
  (vtm 10).
\newblock Technical Report document JVET-S2002, JVET, 2020.

\bibitem{kingma2014adam}
{Adam: A method for stochastic optimization}, author={Kingma, D.P. and Ba, J.}
\newblock {\em arXiv preprint arXiv:1412.6980}, 2014.

\bibitem{frugallydeep2018}
Tobias Hermann.
\newblock {Frugally-Deep}.
\newblock \url{https://github.com/Dobiasd/frugally-deep}, 2018.

\bibitem{bjontegaard2001calculation}
G.~Bjontegaard.
\newblock Calculation of average {PSNR} differences between rd-curves.
\newblock {\em VCEG-M33}, 2001.

\bibitem{jvet_ctc}
J~Boyce, K~Suehring, X~Li, and V~Seregin.
\newblock Jvet common test conditions and software reference configurations.
\newblock Technical Report document JVET-J1010, JVET, 07 2018.

\bibitem{vtmcomplexity}
Gary Sullivan.
\newblock {Versatile Video Coding (VVC) Delivers: Coding Efficiency and Beyond,
  DCC}, 2021.

\bibitem{vvenc}
A.~Wieckowski, J.~Brandenburg, T.~Hinz, C.~Bartnik, V.~George, G.~Hege,
  C.~Helmrich, A.~Henkel, C.~Lehmann, C.~Stoffers, I.~Zupancic, B.~Bross, and
  D.~Marpe.
\newblock Vvenc: An open and optimized vvc encoder implementation.
\newblock In {\em 2021 IEEE International Conference on Multimedia \& Expo
  Workshops (ICMEW)}, pages 1--2, 2021.

\bibitem{vvencai}
I.~Taabane, D.~Menard, A.~Mansouri, and A.~Ahaitouf.
\newblock Machine learning based fast qtmtt partitioning strategy for vvenc
  encoder in intra coding.
\newblock {\em Electronics}, 12(6), 2023.

\bibitem{li2021deepqtmt}
T.~Li, M.~Xu, R.~Tang, Y.~Chen, and Q.~Xing.
\newblock Deepqtmt: A deep learning approach for fast qtmt-based cu partition
  of intra-mode vvc.
\newblock {\em IEEE Transactions on Image Processing}, 30:5377--5390, 2021.

\bibitem{liu2022icip}
Y~Liu, M~Abdoli, T~Guionnet, C~Guillemot, and A~Roumy.
\newblock Statistical analysis of inter coding in vvc test model (vtm).
\newblock In {\em 2022 IEEE International Conference on Image Processing
  (ICIP)}, pages 3456--3459, 2022.

\end{thebibliography}

\ifCLASSOPTIONcaptionsoff
  \newpage
\fi


\vspace{-1.7\baselineskip}

\begin{IEEEbiography}
[{\includegraphics[width=1in,height=1.25in,clip,keepaspectratio]{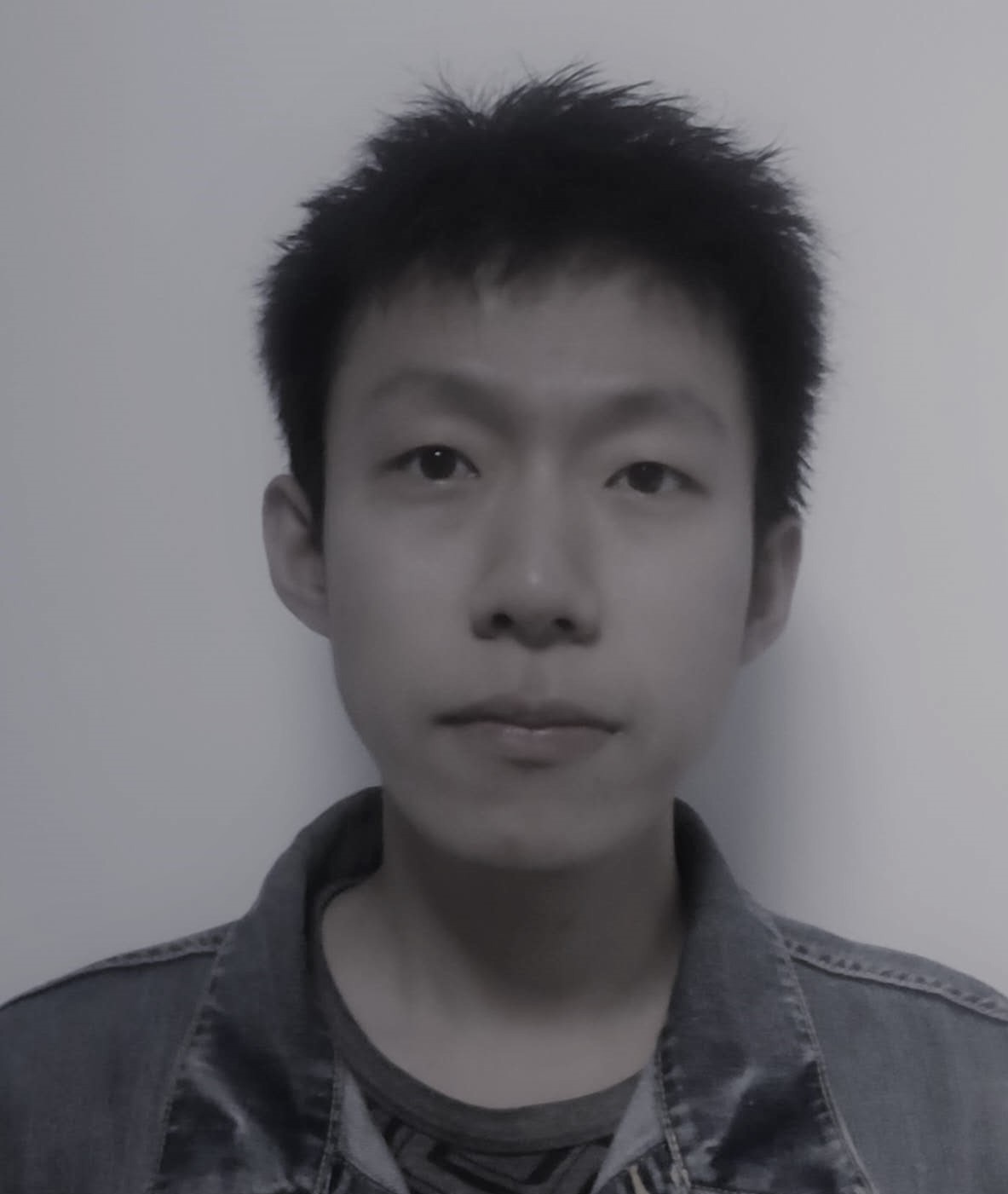}}]{Yiqun Liu}
received the Engineering degree in
electronic information engineering from the Institut national des sciences appliquées de Rennes (INSA Rennes), France, in 2019. He is currently pursuing the Ph.D. degree in partnership
between the Institut National de Recherche en Informatique et en Automatique (INRIA) and the ATEME company, France. His research interests concern video compression and the real-time implementations of the new generation video coding standards.

\end{IEEEbiography}

\vskip -2\baselineskip plus -1fil

\begin{IEEEbiography}
[{\includegraphics[width=1in,height=1.25in,clip,keepaspectratio]{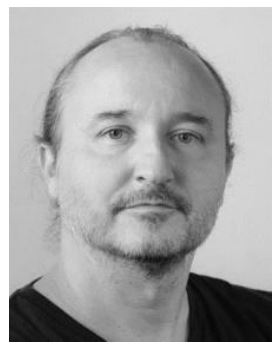}}]{Thomas Guionnet}
is a fellow research engineer at ATEME, where he currently leads the innovation team’s research on artificial intelligence applied to video compression. Beyond his work for ATEME, he has also contributed to the ISO/MPEG - ITU-T/VCEG - VVC, HEVC, and HEVC-3D
standardization process; he teaches video compression at the ESIR Engineering School, Rennes, France; and he has authored numerous publications including patents, international conference papers, and journal articles. Prior to joining ATEME, he spent 10 years at Envivio conducting research on real-time encoding, video-preprocessing, and video quality assessment. He holds a PhD from Rennes 1 University, Rennes.
\end{IEEEbiography}

\vskip -2\baselineskip plus -1fil

\begin{IEEEbiography}
[{\includegraphics[width=1in,height=1.25in,clip,keepaspectratio]{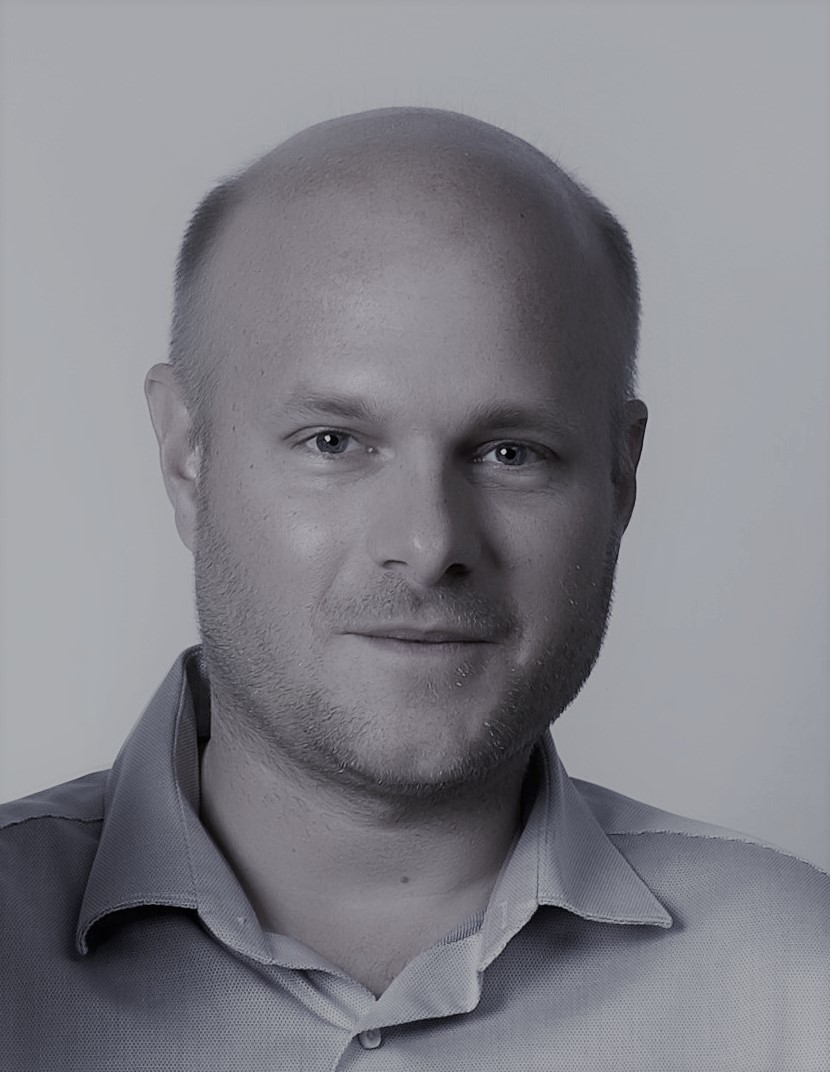}}]{Marc Riviere}
received the MSc degree from the École Nationale de la Statistique et de l'Analyse de l'Information (ENSAI) with highest honors and the MSc degree from the École Nationale Supérieure de Physique de Strasbourg (ENSPS). He joined the CTO Office of ATEME in 2022 as a Senior Data Scientist, applying Artificial Intelligence to various tasks including video analysis, video compression and video distribution. Prior to that, he was an Innovation Architect at the CTO Office of Technicolor Connected Home from 2019 to 2022, where he has notably designed and promoted a large scale data analysis platform. From 2004 to 2018 and within several companies, he has accumulated significant experience in very different topics such as robotics, embedded software, cybersecurity, video watermarking, trusted execution environments, software defined networks and network function virtualization. His current research interests include the areas of machine and deep learning, video compression, computer vision and automatic speech recognition.
\end{IEEEbiography}

\vskip -2\baselineskip plus -1fil

\begin{IEEEbiography}
[{\includegraphics[width=1in,height=1.25in,clip,keepaspectratio]{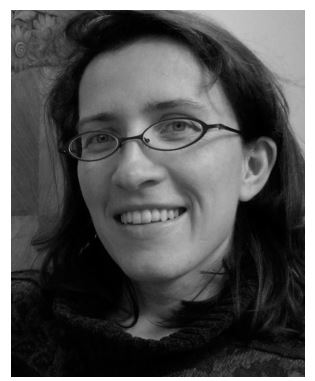}}]{Aline Roumy}
received the Engineering degree from École Nationale Supérieure de l’Éléctronique et de ses Applications (ENSEA), Cergy, France, in 1996, the Master's degree in June 1997, and the Ph.D. degree from the University of Cergy-Pontoise, France, in September 2000. From 2000 to 2001, she was a Research Associate at Princeton University, Princeton, NJ, USA. In November 2001, she joined INRIA Rennes, France. Her current research and study interests include areas of statistical signal
processing, coding theory, and information theory. She was a recipient of a French Defense DGA/DRET Postdoctoral Fellowship from 2000 to 2001.
\end{IEEEbiography}

\vskip -2\baselineskip plus -1fil

\begin{IEEEbiography}[{\includegraphics[width=1in,height=1.25in,clip,keepaspectratio]{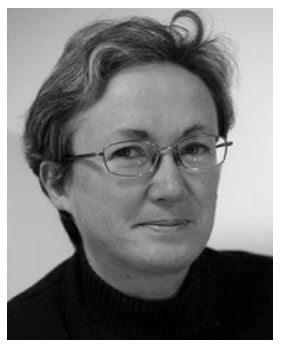}}]{Christine Guillemot, IEEE Fellow}
received the Ph.D. degree from
the École Nationale Supérieure des Télécommunications (ENST), Paris, and the Habilitation degree in research direction from the University of Rennes. From 1985 to 1997, she was with France Télécom, where she was involved in various projects in the area of image and video coding for TV, HDTV, and multimedia. From 1990 to 1991, she was a Visiting Scientist with Bellcore, NJ, USA. She is currently Director of Research with the Institut National de Recherche en Informatique et en Automatique (INRIA) and the Head of a research team dealing with image and video modeling, processing, coding, and communication. Her research interests are signal and image processing, and, in particular, 2D and 3D image and video processing for various problems, such as compression, and inverse pproblems such as restoration, super-resolution, inpainting. Dr. Guillemot served as a Senior Member of the Editorial Board of the IEEE JOURNAL OF SELECTED TOPICS IN SIGNAL PROCESSING from 2013 to 2015. She has served as an Associate Editor for IEEE TRANSACTIONS ON IMAGE PROCESSING from 2000 to 2003 and from 2014 to 2016, IEEE TRANSACTIONS ON CIRCUITS AND SYSTEMS FOR VIDEO TECHNOLOGY from 2004 to 2006, and IEEE TRANSACTIONS ON SIGNAL PROCESSING from 2007 to 2009. She has also been Senior Area Editor of the IEEE TRANSACTIONS ON IMAGE PROCESSING (2016–2020).
\end{IEEEbiography}

\enlargethispage{-5in}

\end{document}